\documentclass[aps,pre,onecolumn,superscriptaddress,10pt,floatfix]{revtex4}

\usepackage[dvips]{graphics}
\usepackage{graphicx}
\usepackage{graphics}
\usepackage{amsfonts}
\usepackage{amssymb}
\usepackage{amsmath}
\usepackage{subfigure}
\usepackage{epstopdf}

\usepackage{tikz}
\usepackage{circuitikz}

\DeclareGraphicsExtensions{.pdf,.eps,.png,.jpg,.mps}
\DeclareMathOperator{\sech}{sech}

\newcommand{\w}{\omega}
\newcommand{\ds}{\displaystyle}
\newcommand{\beq}[0]{\begin{equation}}
\newcommand{\eeq}[0]{\end{equation}}

\begin{document}
\title{Bright Breathers in Nonlinear Left-Handed Metamaterial Lattices}

\author{V.\ Koukouloyannis}
\affiliation{Department of Mathematics and Statistics, University of Massachusetts, Amherst MA 01003-4515, USA}
\affiliation{Department of Mathematics, Statistics and Physics, College of Arts and Sciences, Qatar University, P.O. Box 2713, Doha, Qatar}
\affiliation{Faculty of Civil Engineering, School of Engineering, Aristotle University of Thessaloniki, 54249, Thessaloniki, Greece}

\author{P. G.\ Kevrekidis}
\affiliation{Department of Mathematics and Statistics, University of Massachusetts, Amherst MA 01003-4515, USA}

\author{G. P. Veldes} 
\affiliation{Department of Physics, National and Kapodistrian University of Athens, Panepistimiopolis, Zografos, Athens 15784, Greece}
\affiliation{Department of Electronics Engineering, Technological Educational Institute of Central Greece, Lamia 35100, Greece}

\author{D. J.\ Frantzeskakis}
\affiliation{Department of Physics, National and Kapodistrian University of Athens, Panepistimiopolis, Zografos, Athens 15784, Greece}

\author{D. DiMarzio}
\affiliation{NG Next, Northrop Grumman Corporation, One Space Park, Redondo Beach, CA 90278 USA}

\author{X. Lan  }
\affiliation{NG Next, Northrop Grumman Corporation, One Space Park, Redondo Beach, CA 90278 USA}

\author{V. Radisic}
\affiliation{NG Next, Northrop Grumman Corporation, One Space Park, Redondo Beach, CA 90278 USA}

\begin{abstract}
In the present work, we examine a prototypical model for the
formation of bright breathers in nonlinear left-handed
metamaterial lattices. Utilizing the paradigm of nonlinear
transmission lines, we build a relevant lattice and develop
a quasi-continuum multiscale approximation that enables us to appreciate
both the underlying linear dispersion relation and the potential
for bifurcation of nonlinear states. We focus here, more specifically,
on bright discrete breathers which bifurcate from the lower edge
of the linear dispersion relation at wavenumber $k=\pi$. Guided
by the multiscale analysis, 
we calculate numerically both the stable inter-site 
centered and the unstable site-centered members of the relevant family.
We quantify the associated stability via Floquet analysis and the
Peierls-Nabarro barrier of the energy difference between these
branches. Finally, we explore the dynamical implications of these
findings towards the potential mobility or lack thereof (pinning)
of such breather solutions.
\end{abstract}

\maketitle

\section{Introduction}

A major thrust of theoretical and experimental studies
within electromagnetic media in the last fifteen years
has been allocated to the study of artificially engineered, 
so-called metamaterial structures~\cite{review1,review2,review3}.
Among them, left-handed metamaterials (LHMs) feature
(in specific frequency bands)
simultaneously negative effective permittivity
and permeability, still resulting in a real propagation speed.
The consequent feature of opposite directions of transmission
of energy and wavefronts (or of antiparallel group and phase
velocities) has been examined at microwave, as
well as optical frequencies~\cite{exp1,exp2,shalaev}.

On the other hand, from the nonlinear science perspective,
the same time frame has seen an explosion of interest
in structures that are time-periodic and exponentially localized
in space, the so-called discrete breathers~\cite{aubry,FlachPR2008}.
The experimental relevance of these entities in contributing 
to the localization of energy has been recognized in a broad 
and diverse array of settings. These range from 
micro-mechanical cantilever arrays \cite{SatoPRL2003,SatoC2003},
and  torsionally-coupled pendula~\cite{cuevas} 
to electrical transmission lines \cite{EnglishPRE2008}. They also extend
to Josephson junction superconducting arrays~\cite{TriasPRL2000,BinderPRL2000}, 
coupled antiferromagnetic layers \cite{SchwarzPRL1999}, 
halide-bridged transition metal complexes~\cite{swanson} and, more recently, 
to material science themes such as granular crystals~\cite{BoechlerPRL2010,jinkyu}. 
It should also be added that discrete breathers have also been predicted to 
occur in nonlinear magnetic metamaterials composed by split-ring resonators (SRRs) 
\cite{tsironis} (see also the review~\cite{tsironis2} and references therein) {as have been other localized structures such as bright and dark solitons \cite{laztsi05,kourlaz07}.}

Our aim in the present work is to explore a prototypical 
case example of the interface between these two booming fields, 
utilizing the framework of transmission line (TL) theory.  
This is a method that has been widely used for the study 
of LHMs~\cite{el,Caloz1}. Here, the direct connection of 
electromagnetic properties is between the effective permittivity 
$\epsilon$ and permeability $\mu$ of the LHM and, respectively, 
the serial and shunt impedance of the TL model. Our particular focus 
will be on the nonlinear LHMs where the dependence of the impedances 
on the voltage (or of the effective permittivity/permeability
properties on the electromagnetic field) generates a wealth of relevant
phenomenology. This has been widely explored via the embedding of
wires and SRRs into nonlinear 
dielectrics~\cite{zharov,agranovich}, as well as via the embedding 
of diodes into elements such as the SRRs~\cite{lapine,soukoulis,ysk}.
A broad array of studies have focused on the TL description and its
ability to feature nonlinear structures; see, e.g., the review 
of~\cite{revtl}. More recent theoretical works have also 
utilized the framework of the nonlinear Schr\"{o}dinger (NLS) equation, 
enabling the description of bright~\cite{ogas} or dark~\cite{wang} solitons, 
which have been observed in experiments.

In our specific example of interest, a TL-analogue of a left-handed medium 
is explored, as an extension of our earlier considerations~\cite{inter}. 
These were, at least in part, motivated by the development of strongly 
nonlinear and voltage symmetric barium strontium titanate (BST) thin film 
capacitors~\cite{Meyers}. This symmetric dependence of the capacitance 
of the left-handed element on the voltage has been appropriately 
incorporated in our analysis. Our aim herein is to go a significant 
step further than previous studies in establishing the existence, 
stability, physical relevance and dynamical properties of discrete 
breathers in such systems. In earlier works, to the best of our 
knowledge, upon development of the relevant TL model, NLS-model 
reductions were used to identify an initial ansatz of an approximate soliton  
solution and subsequently these were used in the original dynamical 
lattice to assess their potential robustness. Here, we do not 
restrict our considerations to that. We use the NLS to assess 
the linear properties and (weakly) nonlinear potential of the model 
to carry suitable excitations. Yet, we then go on to 
numerically calculate the true nonlinear bright 
breather solutions (up to a prescribed numerical accuracy) and assess their spectral stability, using 
Floquet theory. We identify both stable solutions (centered
between adjacent lattice sites) and unstable ones (centered on
a lattice site) and quantify their energy difference (the 
celebrated Peierls-Nabarro (PN) barrier~\cite{PN}). Finally, 
we explore the dynamical implications of our stability conclusions 
by initializing the unstable solutions with a perturbation bearing 
a component along this unstable eigendirection. We observe then, a transition of the solutions from a situation of high mobility 
when the PN barrier is low, to one of low mobility and eventual 
trapping/pinning between two adjacent lattice sites as their frequency (which acts as a control parameter) decreases. 

Our presentation is structured as follows. In section II,
we present the model and its theoretical analysis, including 
the quasi-continuum approximation leading to an approximate NLS 
waveform emulating the breather state. Then, in section III, we 
validate and extend the relevant results via detailed numerical 
computations, identifying the two families of breather solutions, 
exploring their stability and dynamics. Finally, in section IV, 
we summarize our findings and present some conclusions for future work.

\section{Model and Theoretical Analysis}

\subsection{The Model}

Following up on our earlier work~\cite{inter}, we build
a lattice of coupled units, each of which
constitutes a nonlinear nodal element, as shown in Fig.~1.

\begin{figure}[h]
\includegraphics[width=8cm]{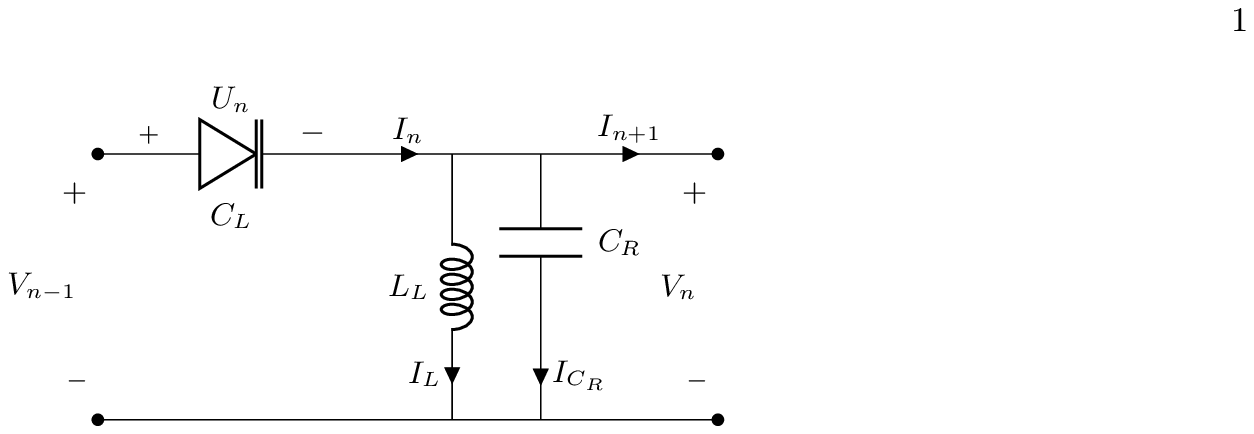}
\caption{A sketch of the unit cell of the transmission line emulating
the left-handed metamaterial.}
\label{expsys}
\end{figure}
Kirchhoff's laws for this unit element circuit read
\begin{eqnarray}
I_{n}&=&I_L+I_{C_R}+I_{n+1}, 
\label{KirchhoffI}\\
U_n&=&V_{n-1}-V_{n}.
\label{KirchhoffIb}
\end{eqnarray}
Taking into regard that 
\begin{eqnarray}
\ds I_{n}=\frac{d}{dt}(C_L(U_n)\,U_n), \quad 
\ds I_{n+1}=\frac{d}{dt}(C_L(U_{n+1})\,U_{n+1}), \quad 
\ds I_{C_R}=\frac{d}{dt}(C_R V_{n}), \nonumber
\end{eqnarray}
substitution to Eqs.~(\ref{KirchhoffI})-(\ref{KirchhoffIb}) yields:
\begin{equation}L_L\frac{d^2}{dt^2}\left[C_L(U_n)\,U_n - C_L(U_{n+1})\,U_{n+1}\right]-L_LC_R\frac{d^2}{dt^2}\left(V_n\right)-V_n=0,
\label{eqU}
\end{equation}
where we have used also $\ds L_L\frac{d I_L}{dt}=V_n$. 
We have thus expressed the dynamical model 
as a second-order dynamical equation modeling the voltages along the TL.

Motivated by the barium strontium titanate
(BST) thin film capacitors of~\cite{Meyers} which feature a strongly
nonlinear symmetric dependence on the voltage,
we explore a dependence of $C_L$ from $U_L$ which is  symmetric
i.e., $C(U)=C_0+\alpha U^2$. Thus, Eq.~(\ref{eqU}) becomes:  
$$L_LC_0\frac{d^2}{dt^2}\left[V_{n-1}-2V_n+V_{n+1}\right]-L_LC_R\frac{d^2V_n}{dt^2}-V_n+\alpha L_L\frac{d^2}{dt^2}\left[(V_{n-1}-V_n)^3-(V_n-V_{n+1})^3\right]=0.\label{eqV}$$
To express the above equation in dimensionless units, we measure
time in $1/\w_0=\sqrt{L_LC_0}$ and voltage in units
of $\sqrt{C_0/\alpha}$, so that
the above equation becomes
\begin{eqnarray}
\frac{d^2}{dt^2}(V_{n-1}-2V_n+V_{n+1})-g \frac{d^2 V_n}{dt^2}-V_n
+\frac{d^2}{dt^2} [(V_{n-1}-V_n)^3 - (V_n-V_{n+1})^3]=0,
\label{modp}
\end{eqnarray}
where $g=C_R/C_0$. 
In what follows the parameter $g$ 
will play a critical role for 
the existence of bright breathers in this system.

It is now useful, for the purposes of our analytical and
numerical considerations, to adopt experimentally relevant
parameter values. Left-handed metamaterials featuring the equivalent lattice 
model of Fig.~\ref{expsys} can be fabricated in planar structures using lumped 
elements, or distributed components, in microwave frequencies.
In this case, the values of electrical parameters of equivalent lattice model, 
namely the capacitances and inductances can be extracted from the physical 
parameters of these structures~\cite{kong}. For instance, for an array of complementary 
SRRs (CSRRs) \cite{beruete}, depending on the size of these resonant elements, and the type 
of the diode that is inserted in the slits of the CSRRs, one can use 
the following parameter values. 
In the case of relatively small CSRRs, operating in the microwave 
frequency range $1.77$~GHz~$<f<2.7$~GHz, one may use 
a varactor diode MA46H146 \cite{macom}, and obtain: 
$C_0 \approx 0.5$~pF, $L_{L}\approx 2.3$~nH, and $C_{R}\approx 1.5$~pF; 
this choice leads to $g\simeq3$. 
On the other hand, in the case of relatively large CSRRs, operating 
in the frequency range  $130$~MHz~$<f<1.100$~MHz, 
one may use a barium strontium titanate (BST) 
thin film capacitor \cite{Meyers}, and obtain: $C_0\approx80$~pF, 
$L_{L}=4.7$~nH and $C_{R}=4.5$~pF; this results in 
$g\simeq0.056$.

\subsection{Energy of the system}

Here, we will introduce the energy of the system for a number
of both theoretical and practical reasons. On the one hand,
the total energy is a conserved quantity in the system and yields a practical
diagnostic for the accuracy, e.g., of the numerical simulations performed
below. On the other hand, its density is a measure of the
localization of the coherent structures (breathers) that we will
consider in what follows. Hence, the energy density
provides an intrinsic quantity tailored towards monitoring the
space-time dynamical evolution of the system.

Considering from first principles, the energy in the different
capacitive and inductive elements, we can reconstruct
the energy of the unit-cell of the lattice. This is given by
$$E_u=E_{C_L}+E_{C_R}+E_{L_L}.$$
By using the well known expressions for the above mentioned energies and by recalling the specific form of $C_L-U$ dependence, this becomes 
\beq 
E_u=\frac{1}{2}C_0U^2+\frac{3}{4}\alpha U^4+\frac{1}{2}C_RV_n^2+\frac{1}{2}L_LI_L^2.
\label{unitenergy}
\eeq
Measuring the energy in units of $C_0^2/\alpha$, 
(and also time in $1/\w_0=\sqrt{L_LC_0}$ units, voltage in units
of $\sqrt{C_0/\alpha}$, as before and current in units of
$C_0/\sqrt{\alpha L_L}$),
Eq.~(\ref{unitenergy}) in its rescaled dimensionless form becomes:
\beq E_u=\frac{1}{2}U^2+\frac{3}{4}U^4+\frac{g}{2}V_n^2+
\frac{1}{2}I_L^2.\label{unitenergynorm}\eeq
The total (conserved) energy of the lattice is then reconstructed
upon the relevant summation over nodes:
\beq E_{tot}=\sum_n\left\{\frac{1}{2}(V_n-V_{n+1})^2+\frac{3}{4}(V_n-V_{n+1})^4+\frac{g}{2}V_n^2+\frac{1}{2}I_{L_n}^2\right\}.\label{Etot} \eeq
We can also define the ``energy per site'' as
\beq E_{n}=\frac{1}{4}\left[(V_{n-1}-V_{n})^2+(V_n-V_{n+1})^2\right]+\frac{3}{8}\left[(V_{n-1}-V_{n})^4+(V_n-V_{n+1})^4\right]+\frac{g}{2}V_n^2+\frac{1}{2} I_{L_n}^2,\label{En}\eeq
such that $E_{tot}=\sum_nE_n$.

\subsection{The quasi-continuum approximation}
The mathematical model of Eq.~(\ref{modp}) is
less straightforward to tackle directly. It is for
this reason that 
we employ a quasi-continuum approximation~\cite{Rem}.
This affords us an understanding of both the underlying
linear band structure and that of developing the
weakly nonlinear theory, based on the NLS approximation.

More concretely, we seek 
solutions of Eq.~(\ref{modp}) in the form of the asymptotic expansion: 
\begin{equation}
V_n =\sum_{\ell=1}^{+\infty} \epsilon^{\ell} V^{(\ell)}(X,T)e^{i\ell (\omega t - k n)} +{\rm c.c.},
\label{eq:ansatz}
\end{equation}
where ${\rm c.c.}$ stands for complex conjugate, and
the terms in the power series expansion
$V^{(\ell)}$ are unknown envelope functions, 
that depend on the slow variables:
\begin{equation}
X=\epsilon (n-v_g t), \qquad T = \epsilon^2 t,
\label{slow}
\end{equation}
with $v_g$ being the group velocity, as can be found self-consistently 
from the linear dispersion relation (see below).
The scaling selected here is the standard one associated with the NLS model. 
Finally, $\omega$ and $k$ denote the carrier's frequency and wavenumber, respectively, 
and $\epsilon$ is a formal small parameter, characterizing
the amplitude of the solution (and its inverse spatial width). 

The resulting equations from the multiscale expansion
order by order read (cf. also Ref.~\cite{inter}): 
\begin{eqnarray}
O(\epsilon)\ &:& \frac{1}{ \omega^2 }  =  2 + g- 2 \cos k; \label{disp} \\
O(\epsilon ^{2})&:&  {v_g} = - \omega^3 \sin k ;\qquad V^{(2)} = 0;\label{v_g}\\
O(\epsilon ^{3})&:& i \partial_T V^{(1)}+P \partial_{X}^2 V^{(1)} + Q |V^{(1)}|^2 V^{(1)}=0,
\quad P = \frac{\omega^3}{2} (\cos k-3 \omega^2\sin^2 k ), 
\quad Q = -24
\omega^3\sin^4(k/2);
\label{exp}
\\
&& V^{(3)} = \frac{144
\omega^2 (1+2\cos k ) \sin^4(k/2) {V^{(1)}}^3}{1+g-2\cos(3k)}. 
\nonumber \label{v3}
\end{eqnarray}
The first one of these, at $O(\epsilon)$, represents the linear dispersion
relation of the LHM. 
Notice that the dispersion relation (\ref{disp}) suggests that there exist two 
cutoff frequencies, namely an upper one, $\omega_{max} =1/\sqrt{g}$ (corresponding
to $k= 0$), and a lower one, $\omega_{min}=1/\sqrt{g+4}$ (corresponding to $k = \pi$);
notice that the lower cutoff frequency is due to discreteness since, evidently, 
this frequency vanishes in the continuum limit. 

At the next order, the solvability condition yields
the group velocity (the velocity of wavepackets) $v_g=d\omega/dk$, 
which is not only distinct from the phase velocity $v_p = \omega/k$, but also 
carries opposite sign, as per the left-handed nature of the medium; 
this becomes clear by the form of the dispersion relation shown in the left panel 
of Fig.~\ref{PQ56}, which features a negative slope.
Note that, at the second order, the solvability condition leads
to a vanishing contribution $V^{(2)}=0$, as is commonly the case in such
multiscale expansions (under symmetric conditions). 

At the third order, we obtain the NLS 
equation for $V^{(1)}$. Its dispersion and nonlinearity coefficients, $P$ and $Q$ respectively,
depend on the frequency, but the latter is slaved to the wavenumber through the dispersion 
relation. Last, but not least, the third-order reduction/decomposition
of the solution is also derived. This enables us to reconstruct
the solution up to errors of size $\epsilon^4$. Naturally, we
expect this to be a fairly adequate approximation of the solution
within the weakly nonlinear regime.

The product $PQ$ determines the ``nature'' of the
resulting NLS. In the region where the relevant 
quantity is positive, per the standard general theory
of the NLS equation~\cite{sulem,ablowitz,siambook}, the dynamics is
associated with a self-focusing scenario that supports 
bright solitons. On the other hand, when $P Q <0$, then we are in a self-defocusing
regime where, e.g., dark solitons may arise. The analytical availability
of these waveforms at the NLS level, as well as the explicit form
of the transformation, allows us to express these potential solutions
in the LHM dynamics. 

\subsection{Identifying Bright Breathers}

In this work we are, more specifically, interested in exploring
the existence and stability of bright breather-like solutions. 
For this purpose, we are going to use the NLS bright soliton solution as 
initial estimate for the periodic motion finding numerical scheme, 
which is given by Eq~(\ref{bright}). 
\begin{eqnarray}
V^{(1)} = \sqrt{\frac{2P}{Q}}u_0\sech \left(u_0(X-2c|P|T)\right)
\exp[i\left(cX+(u_0^2-c^2)|P|T\right)],
\label{bright}
\end{eqnarray}
where $u_0$ and $c$ are free O$(1)$ parameters setting the
amplitude/inverse width and wavenumber of the soliton, respectively. 

Since we are seeking standing solutions ($u_g=0$), we require the value 
of the wavenumber to be $k=0$ or $k=\pi$, as can be seen by Eq.~(\ref{v_g}).
In addition, since the bright solution is supported in the $k$-parameter range 
where the NLS demonstrates focusing behavior, i.e. $PQ>0$, the appropriate 
choice of the wavenumber is $k=\pi$, as shown in Fig.~\ref{PQ56} where the 
value of the product $PQ$ is depicted for the two values of $g$ discussed in the previous section, $g=0.056$ (in line
with what was used in Ref.~\cite{inter}) and $g=3$.
More generally, it is well known that the nonlinear solutions 
will bifurcate from the edges of the linear band. Near the 
$k=0$ limit, we can observe that the relevant product 
vanishes, rendering a corresponding nonlinear waveform 
bifurcation less promising. However, the focusing 
nature of the equation near $k=\pi$ is promising towards 
attempting the profile of Eq.~(\ref{bright}), in conjunction with the 
expansion of Eq.~(\ref{eq:ansatz}) and the variables of Eq.~(\ref{slow}), 
to identify a bright breather profile. We will thus utilize this 
selection as an initial guess in our attempt to 
identify numerically exact bright breather solutions 
for frequencies below those of the linear spectrum 
lower band edge in what follows.

\section{Numerical Results}

\subsection{Existence and linear stability of breather families}

\begin{figure}[!h]
\centering
\includegraphics[height=6cm]{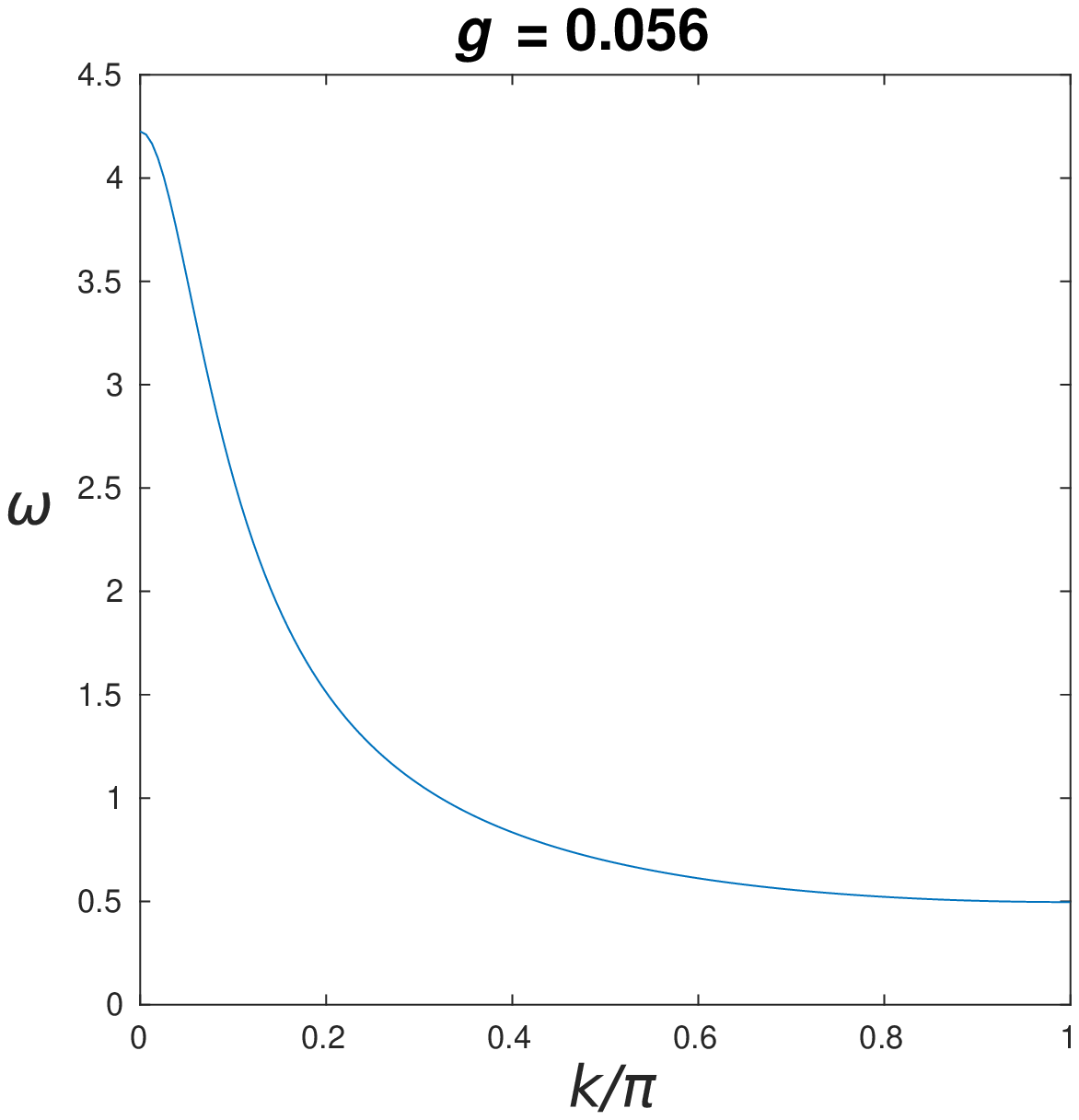}\includegraphics[height=6cm]{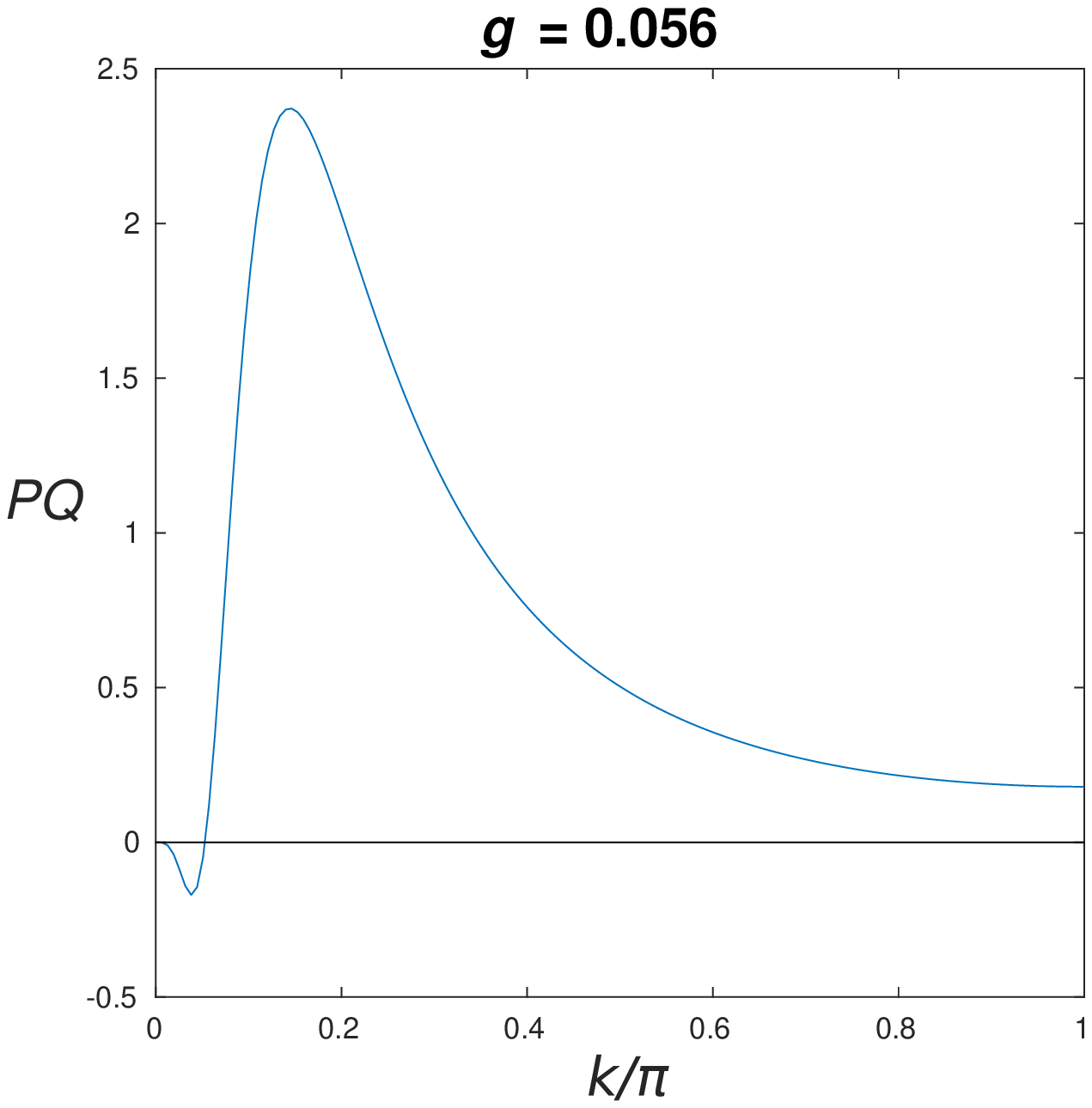}
\includegraphics[height=6cm]{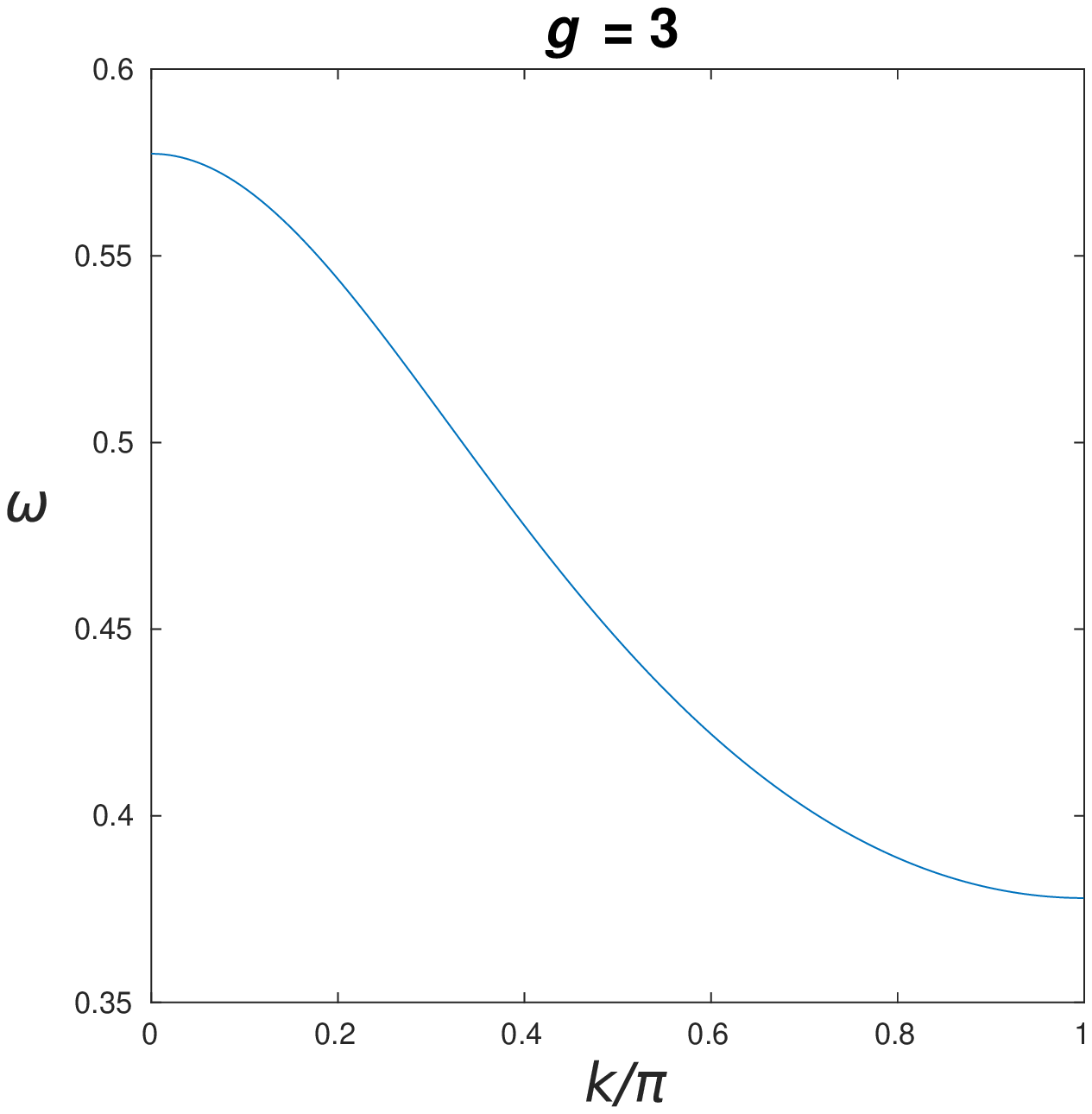}\includegraphics[height=6cm]{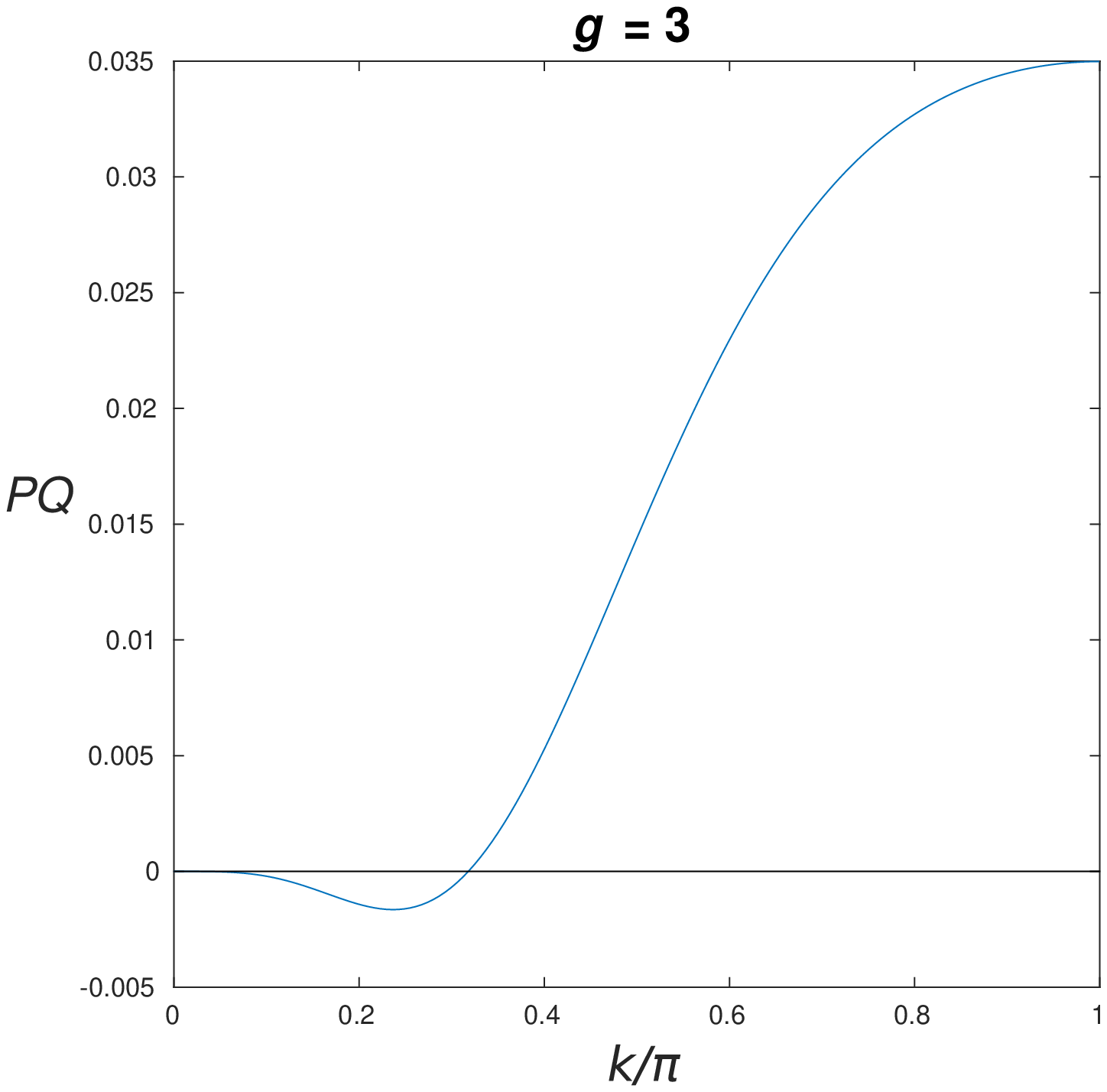}
\caption{(Color online) Top left panel: the linear dispersion relation for $g=0.056$ 
[cf. Eq.~(\ref{disp})]. Top right panel: the dependence of the factor $P Q$ (which determines  
the focusing or defocusing nature of the model) on the frequency $\omega$; 
see also the text. The bottom panels show the same quantities but now for $g=3$.
When $P Q >0$, the nonlinearity is 
self-focusing while for the opposite sign, it is self-defocusing.} \label{PQ56}
\end{figure}

In order for a breather-like solution to exist it is necessary for its frequency $\w=2\pi/T$ 
to lie outside the range of the dispersion relation (see e.g. fig.~\ref{PQ56}). 
In addition, every multiple of this frequency should also avoid this range, 
in order for the conditions of the theorem for the existence of breathers to 
be satisfied~\cite{MA94}. Since this solution will bifurcate from the right end 
of this curve it will lie bellow the curve. As we can easily verify from the top panels 
of Fig.~\ref{PQ56}, this in not possible for $g=0.056$ 
since the second harmonic of 
every choice of frequency below the dispersion relation curve will resonate
with the 
linear spectrum for this choice of $g$. We should note here that in a chain with a few 
nodes, due to the quantization of the wavenumbers (and the resulting ``gaps'' within 
the pass band), these resonances can be avoided. Nevertheless, we do not consider 
this scenario here, having in mind, in principle, the case of an extended 
(practically infinite) lattice. 

In order to manage to avoid the linear spectrum represented by the dispersion 
relation of Eq.~(\ref{disp}), we have to use a value of $g$ that provides 
a suitably narrow linear band. The critical value for which this is possible is $g_{cr}=4/3$. So, 
we choose to use the value of $g=3$ (which is also physically relevant as discussed also in the previous section). 
The corresponding diagrams are shown in the bottom panels of Fig.~\ref{PQ56}. 
The limiting values of $\w$ are $\w_{min}=0.378$ and $\w_{max}=0.577$. 
Thus, if we consider, e.g., the limiting value $\w=0.378$ for the frequency 
of our breather, then the second harmonic $2\w=0.756$ lies above the value 
of $\w_{max}$ and the conditions for the persistence of the breathers 
are automatically satisfied~\cite{MA94}.

\begin{figure}
\centering
\begin{tabular}{cccc}
\includegraphics[scale=0.3]{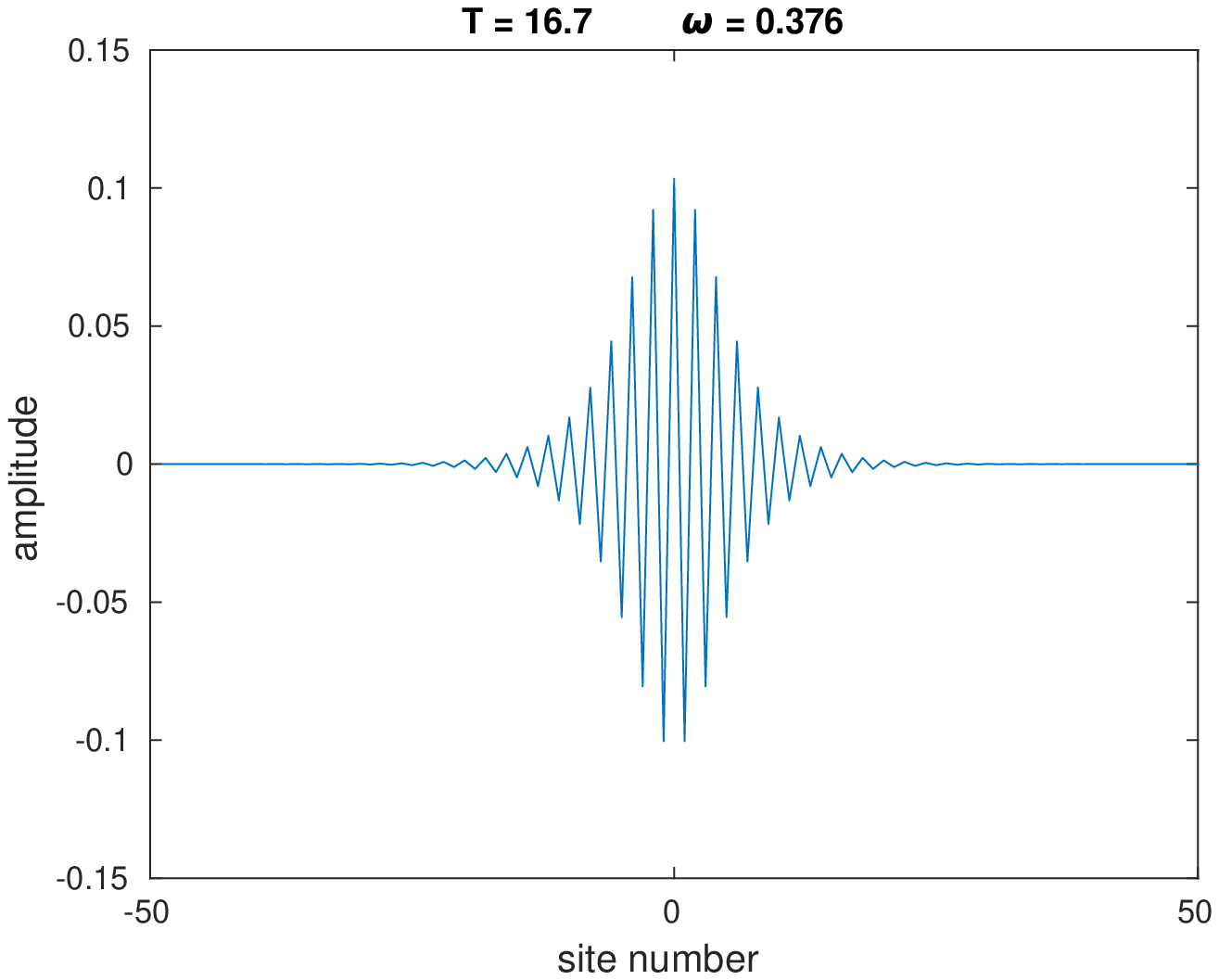}
&\hspace{-0.5cm}
\includegraphics[scale=0.3]{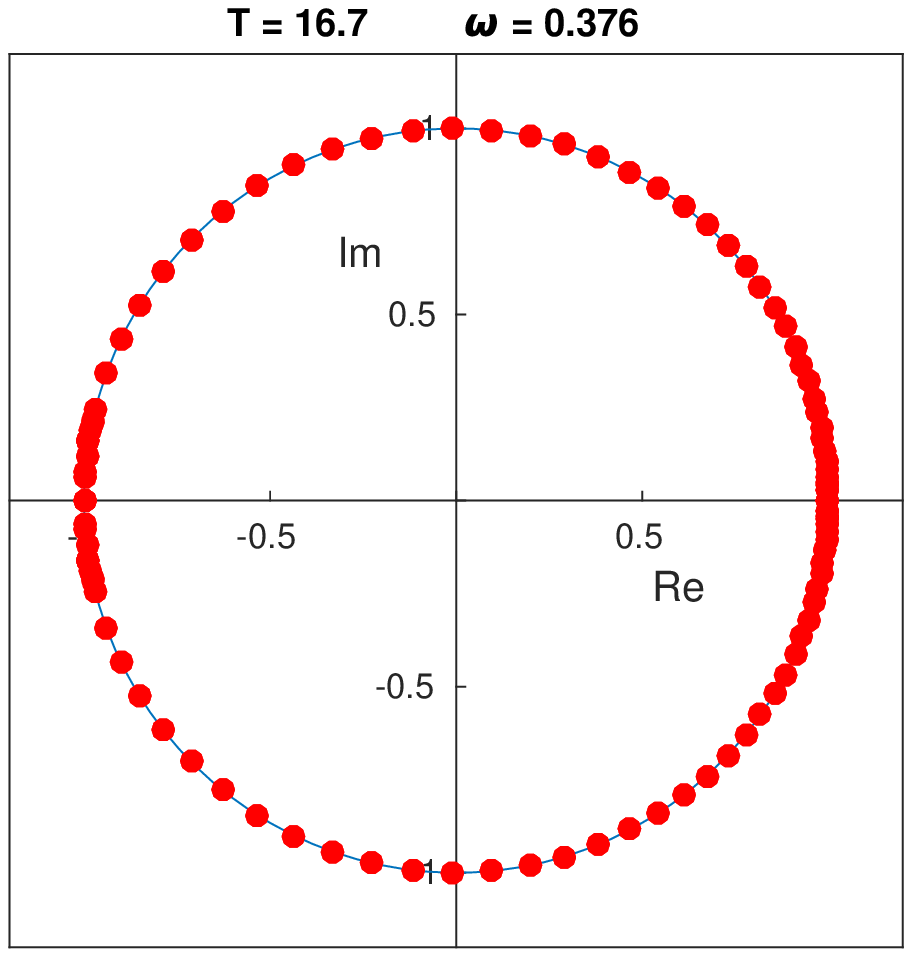}
&\hspace{-0.0cm}
\includegraphics[scale=0.3]{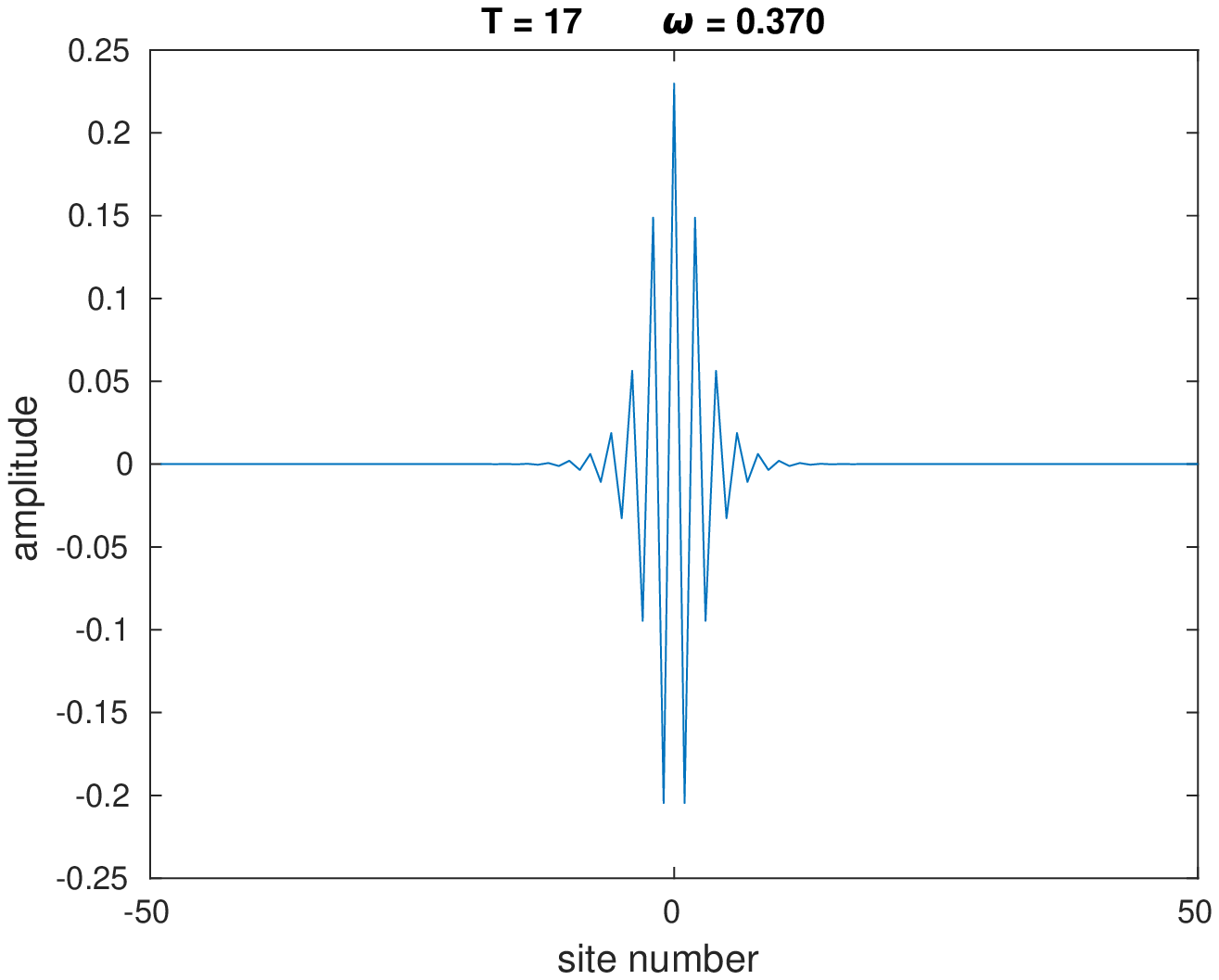}
&\hspace{-0.5cm}
\includegraphics[scale=0.3]{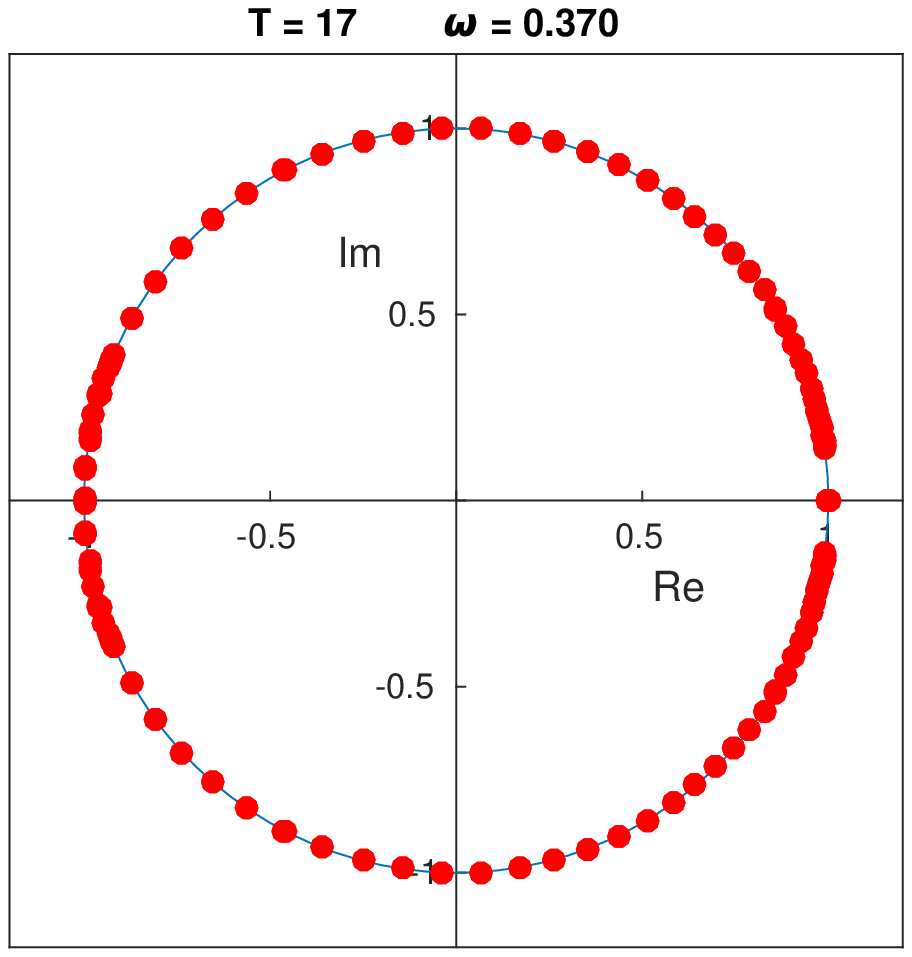}
\\
\includegraphics[scale=0.3]{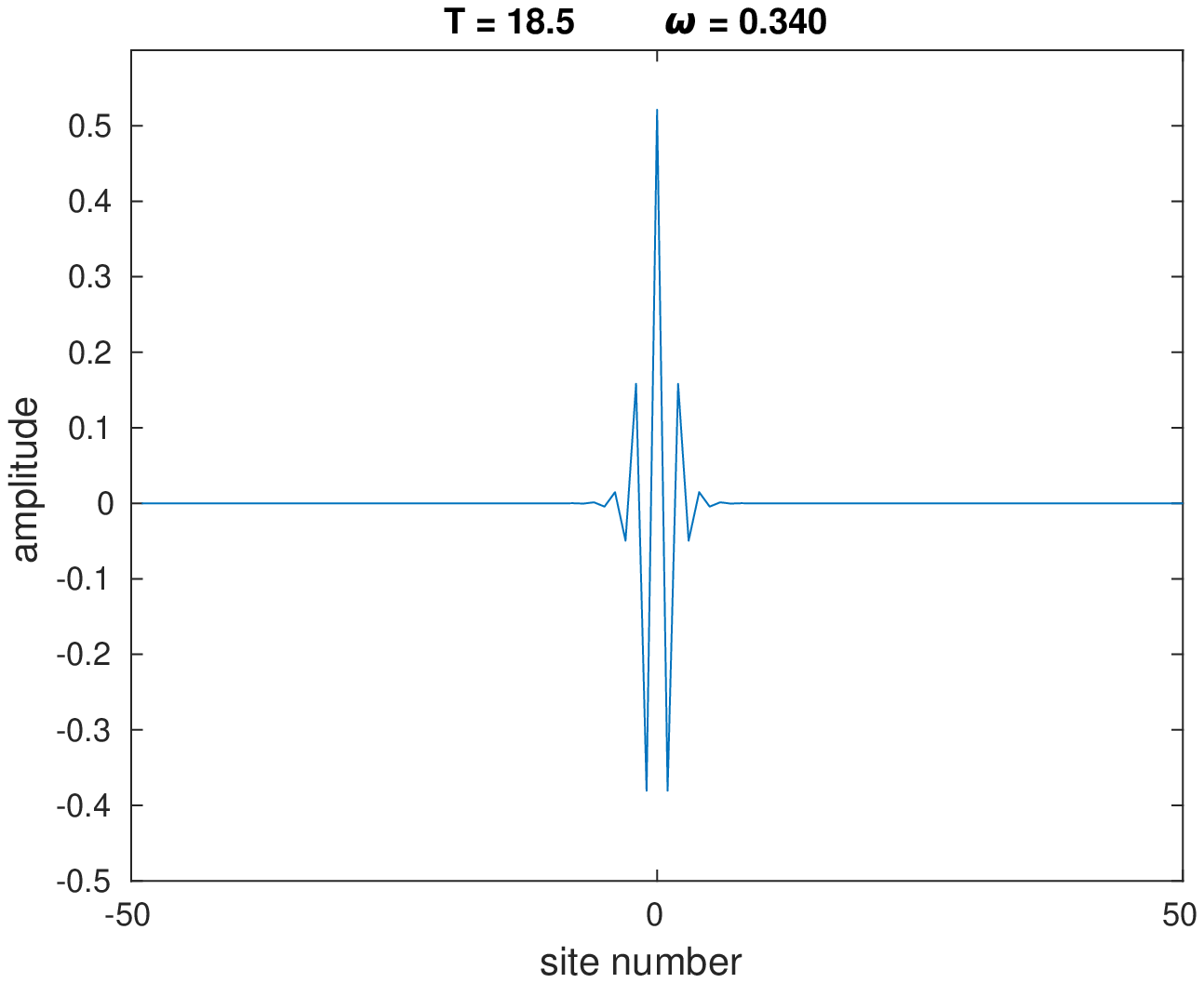}
&\hspace{-0.5cm}
\includegraphics[scale=0.3]{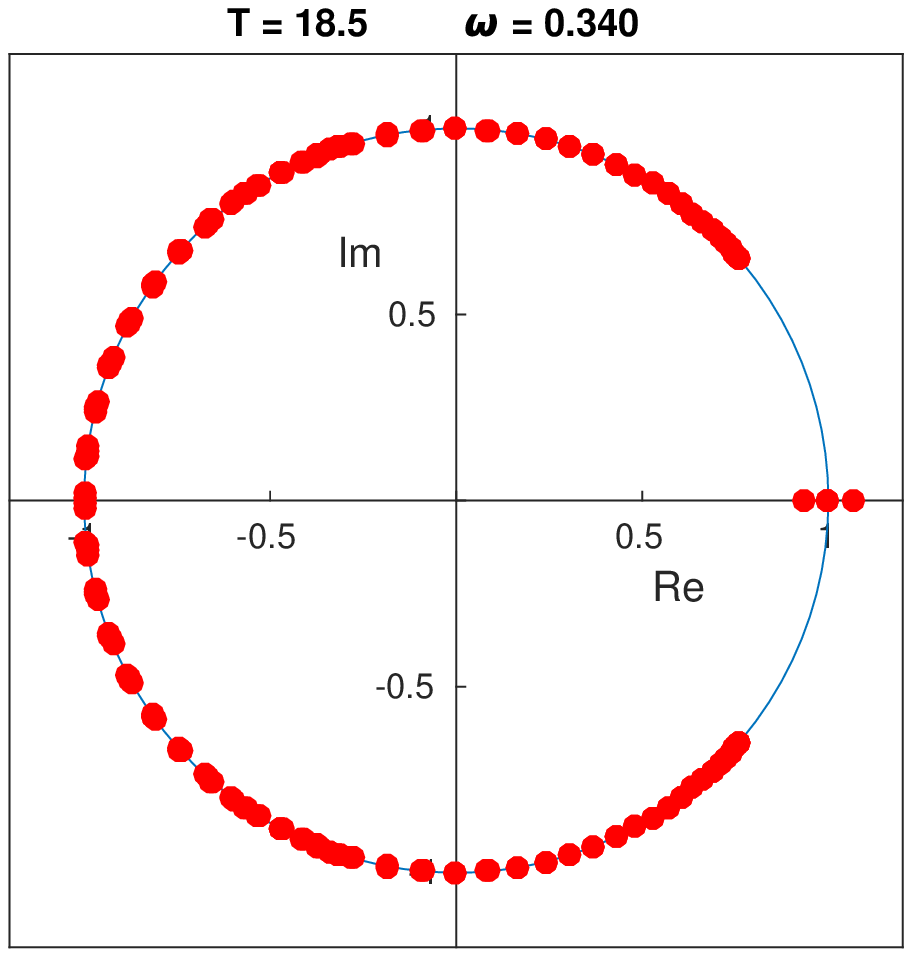}
&\hspace{-0.0cm}
\includegraphics[scale=0.3]{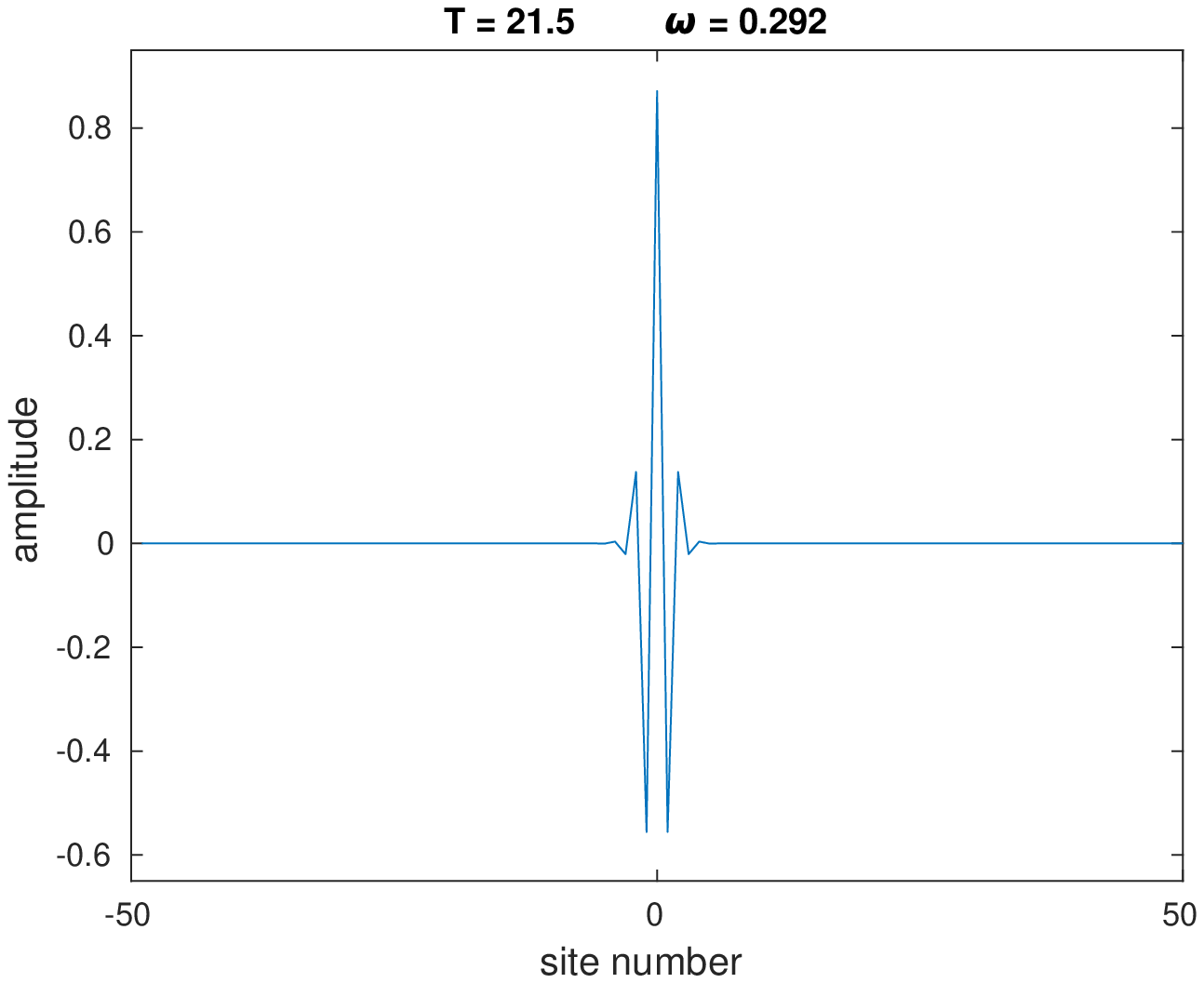}
&\hspace{-0.5cm}
\includegraphics[scale=0.3]{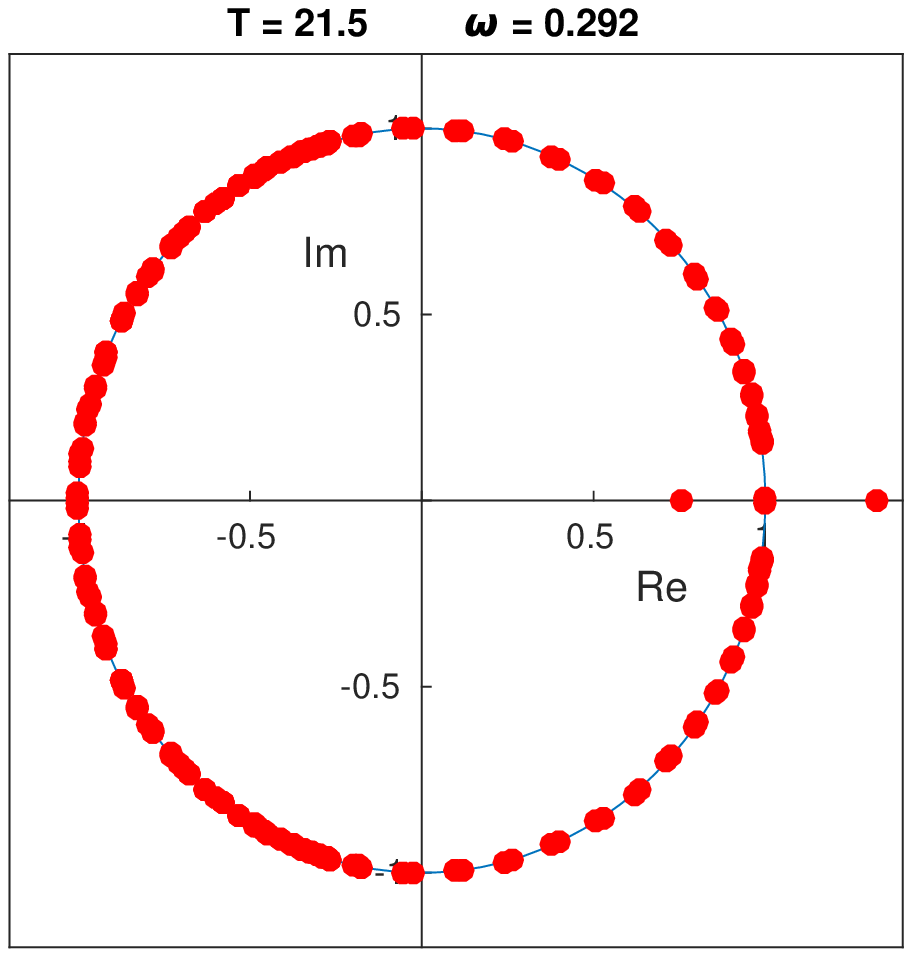}
\\
\end{tabular}
\caption{The onsite family for $T=16.7, 17, 18.5\ \text{and}\ 21.5$ (or equivalently $\w=0.376, 0.370, 0.340\ \text{and}\ 0.292$) together with 
the corresponding Floquet multipiers. We can see here that this family is 
linearly unstable since there is a pair of multipliers leaving the unit circle 
along the real axis.}
\label{onsite}
\end{figure}

The solution of Eq.~(\ref{bright}) for $k=\pi$ is introduced in Eq.~(\ref{eq:ansatz}), 
together with the third-order contribution $V^{(3)}$ of Eq.~(\ref{v3}).
Subsequently, the numerical scheme is iterated leading, upon 
convergence, to the identification of the periodic orbit and the
associated period. For this initial estimate and for $T=16.7\Rightarrow\w=0.376$ 
we get two distinct breather solutions, namely 
the onsite and inter-site one. After acquiring the breathers for
a particular value of the frequency, 
we follow a continuation procedure and get two one-parameter families of breathers 
with $\w$ as the corresponding parameter. The onsite family, 
for $T=16.7$, $17$, $18.5$ and $21.5$ (or equivalently
$\w=0.376$, $0.370$, $0.340$ and $0.292$
is shown in Fig.~\ref{onsite}, 
together with the corresponding Floquet multipiers $\lambda_i$.
Using the monodromy matrix and identifying its eigenvalues, 
as per the well-known numerical technology discussed, e.g., 
in Ref.~\cite{FlachPR2008}, we can identify the stability of these 
solutions. When the $\lambda_i$'s are all on the unit circle, 
the corresponding solutions are stable, while pairs on the 
real axis (exponential growth), or quartets off of the unit 
circle (oscillatory growth) are associated with instability. 
The onsite family is linearly unstable since there is always 
a pair of multipliers leaving the unit circle along the real axis. 
The value of the largest characteristic exponent $\mu_i$ with 
respect to $\w$ is shown in Fig.~\ref{bifdiag}. 

\begin{figure}
\centering
\begin{tabular}{cccc}
\includegraphics[scale=0.3]{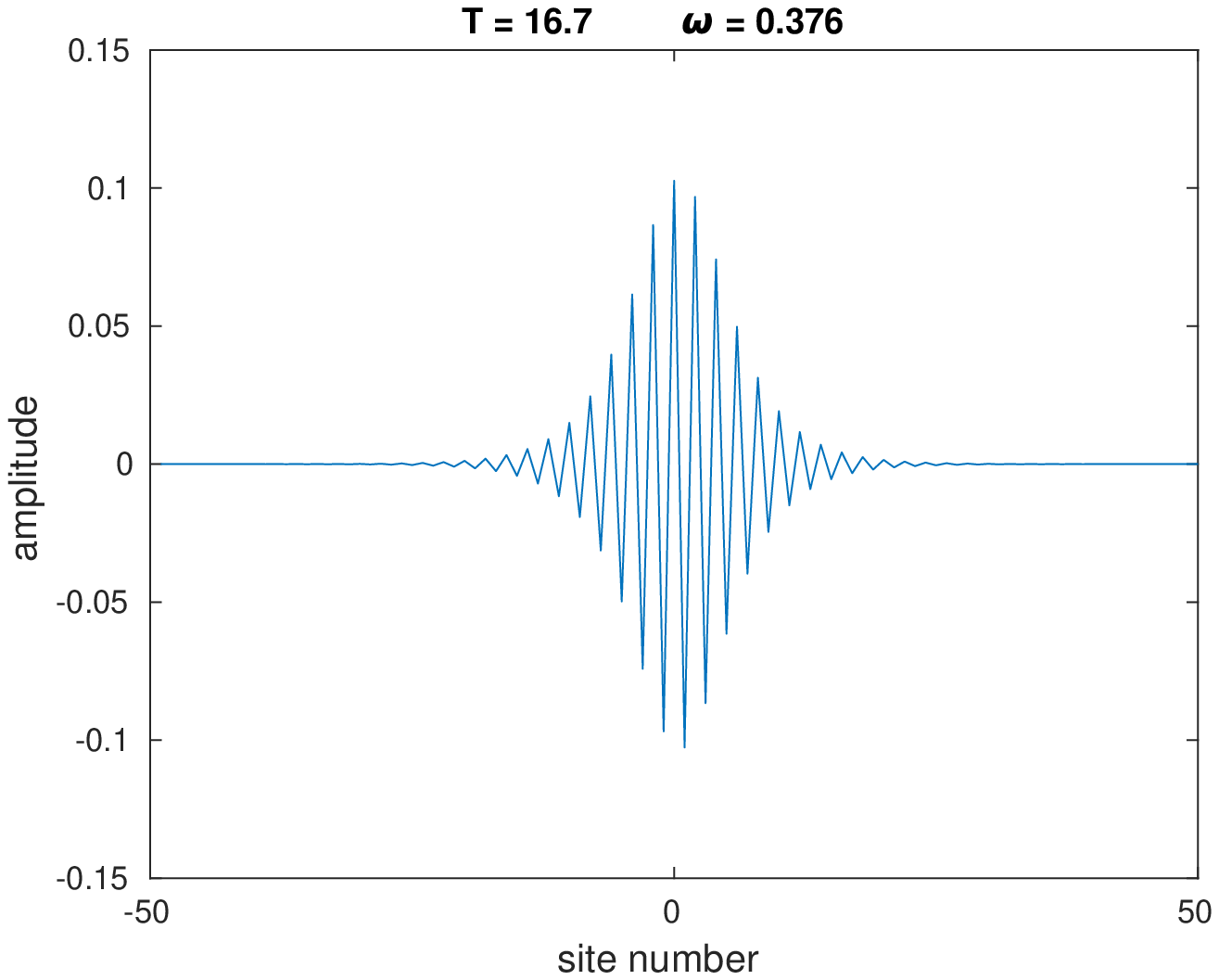}
&\hspace{-0.5cm}
\includegraphics[scale=0.3]{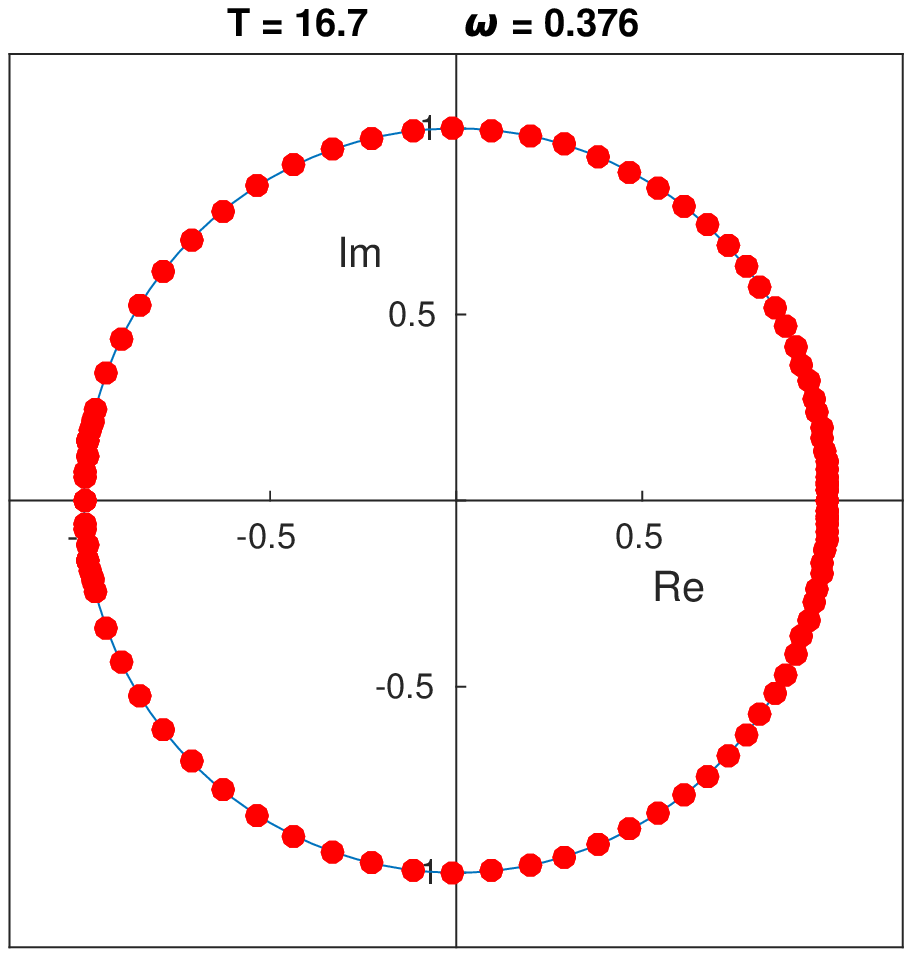}
&\hspace{-0.0cm}
\includegraphics[scale=0.3]{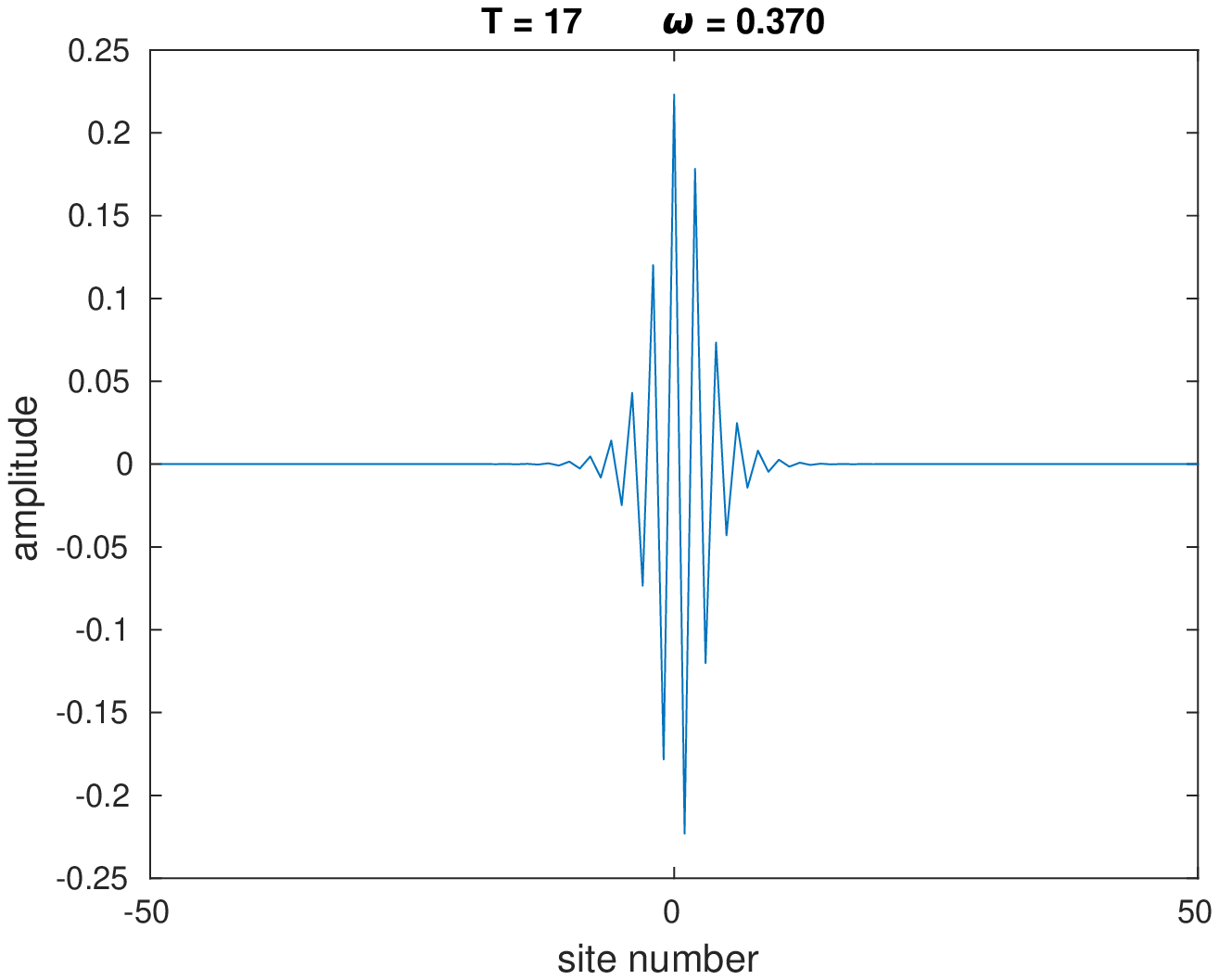}
&\hspace{-0.5cm}
\includegraphics[scale=0.3]{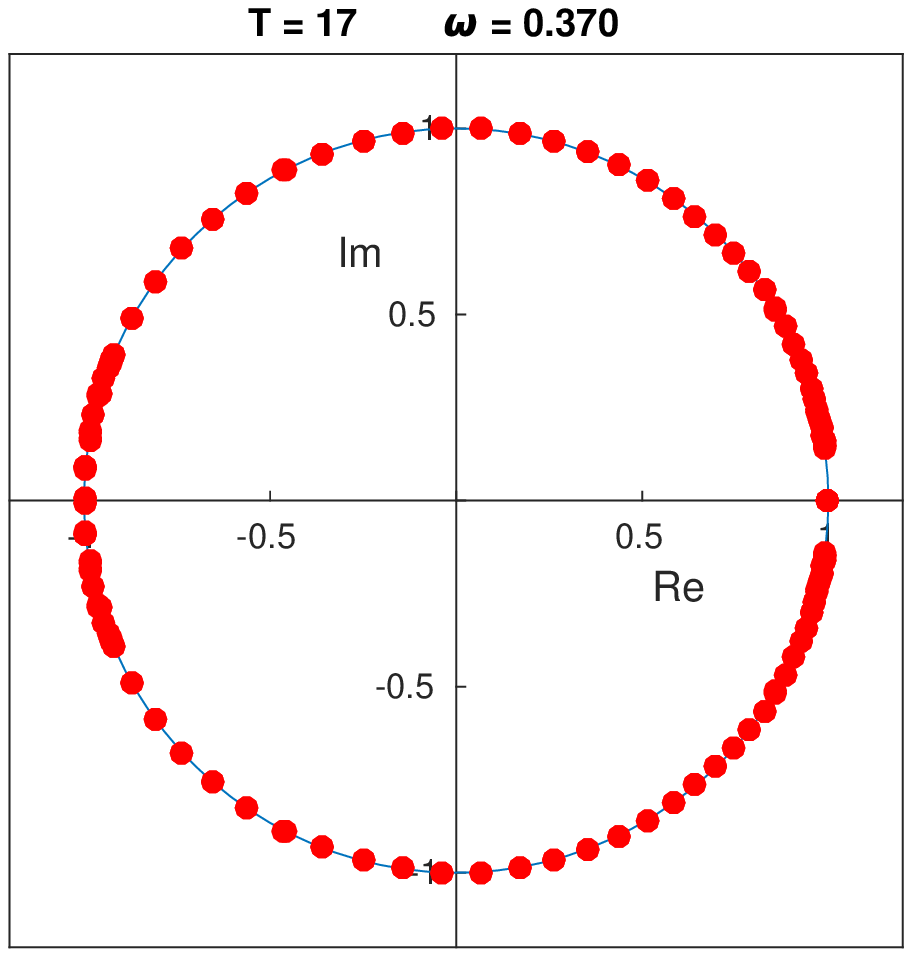}
\\
\includegraphics[scale=0.3]{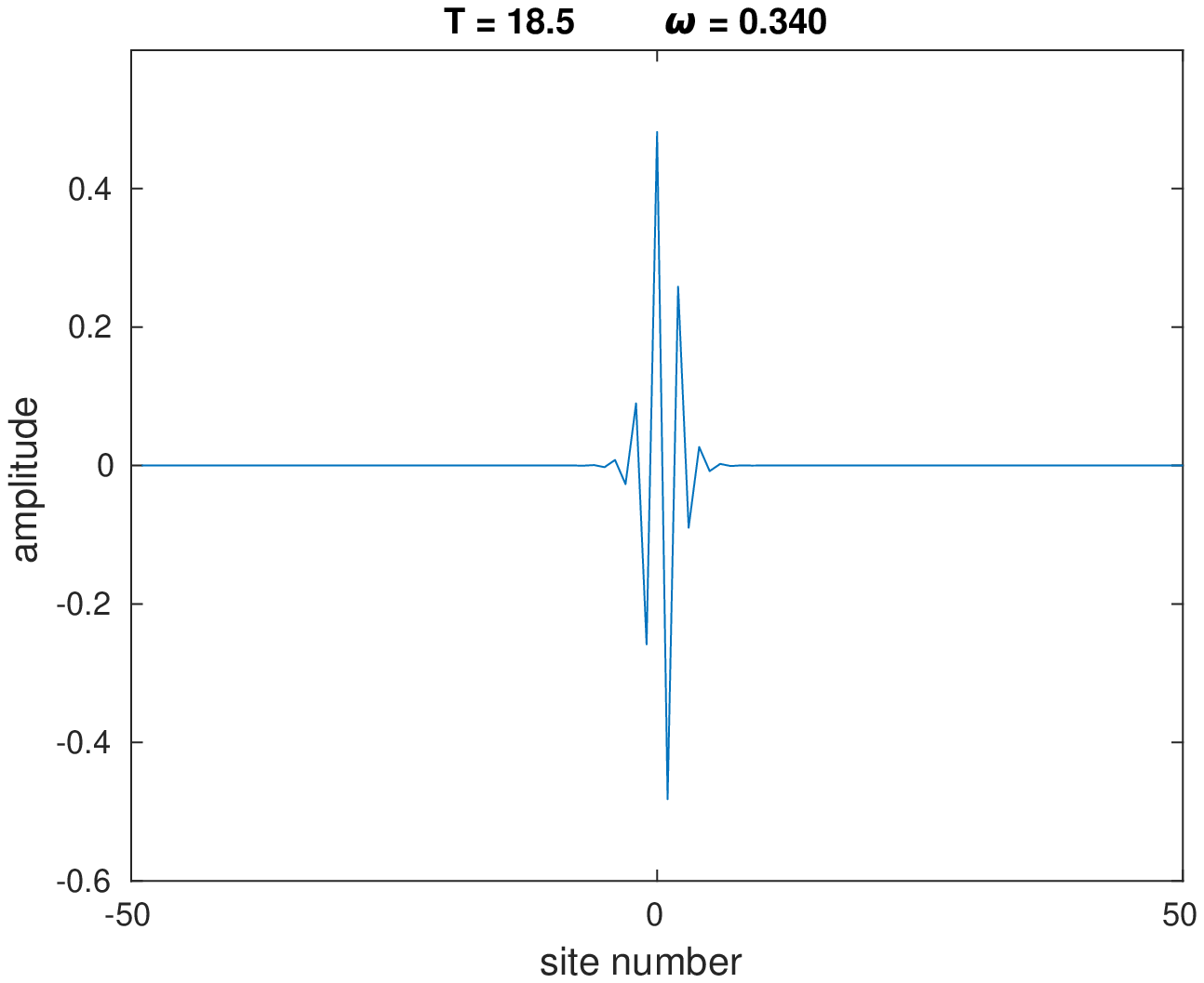}
&\hspace{-0.5cm}
\includegraphics[scale=0.3]{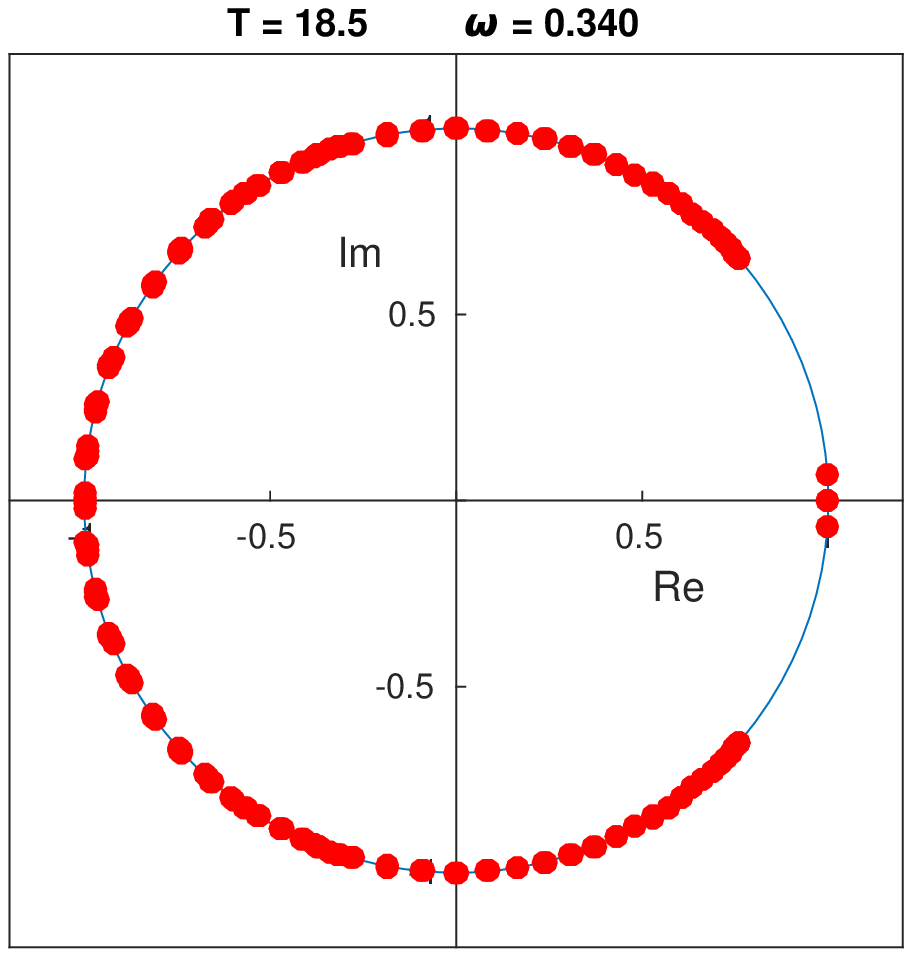}
&\hspace{-0.0cm}
\includegraphics[scale=0.3]{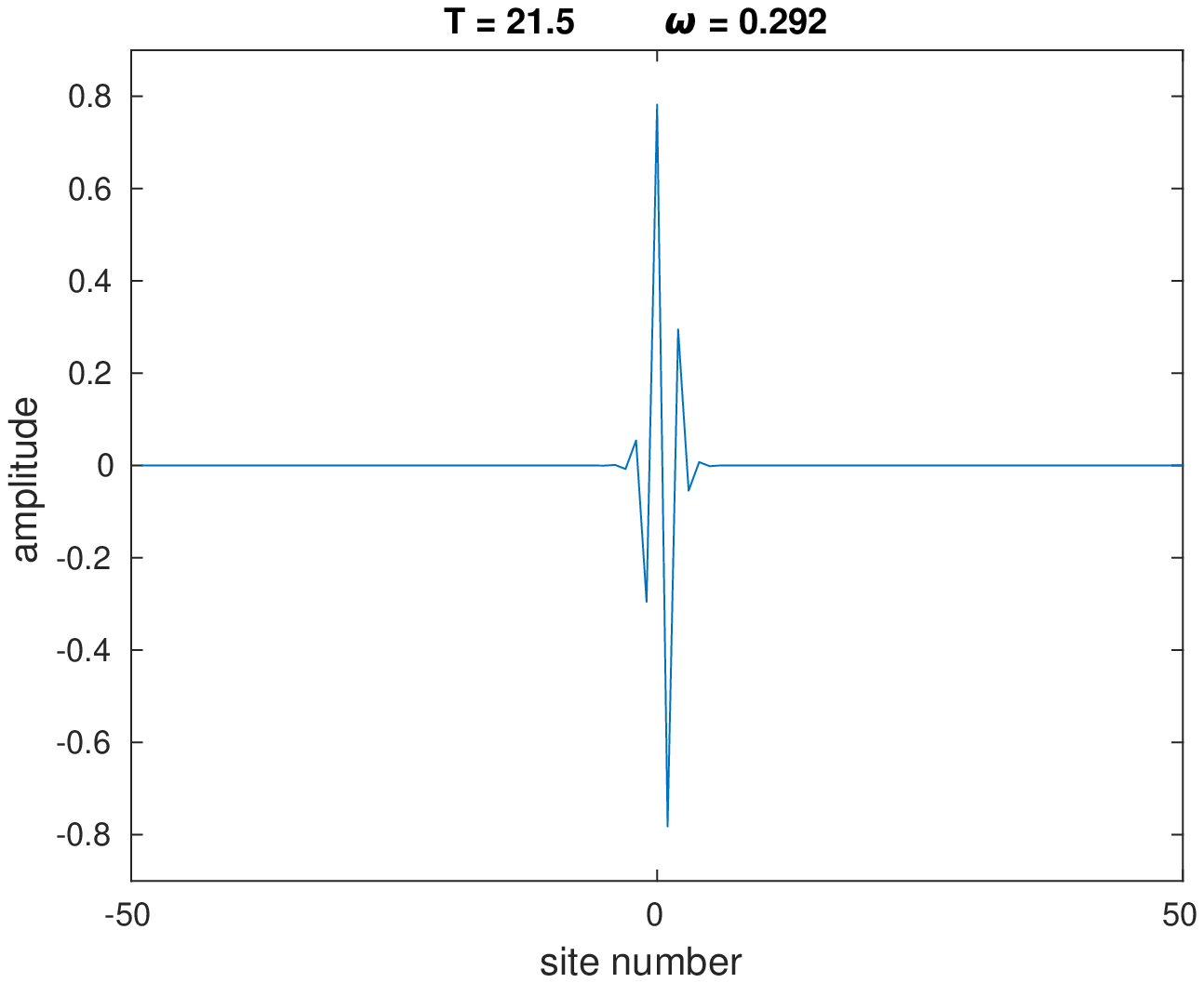}
&\hspace{-0.5cm}
\includegraphics[scale=0.3]{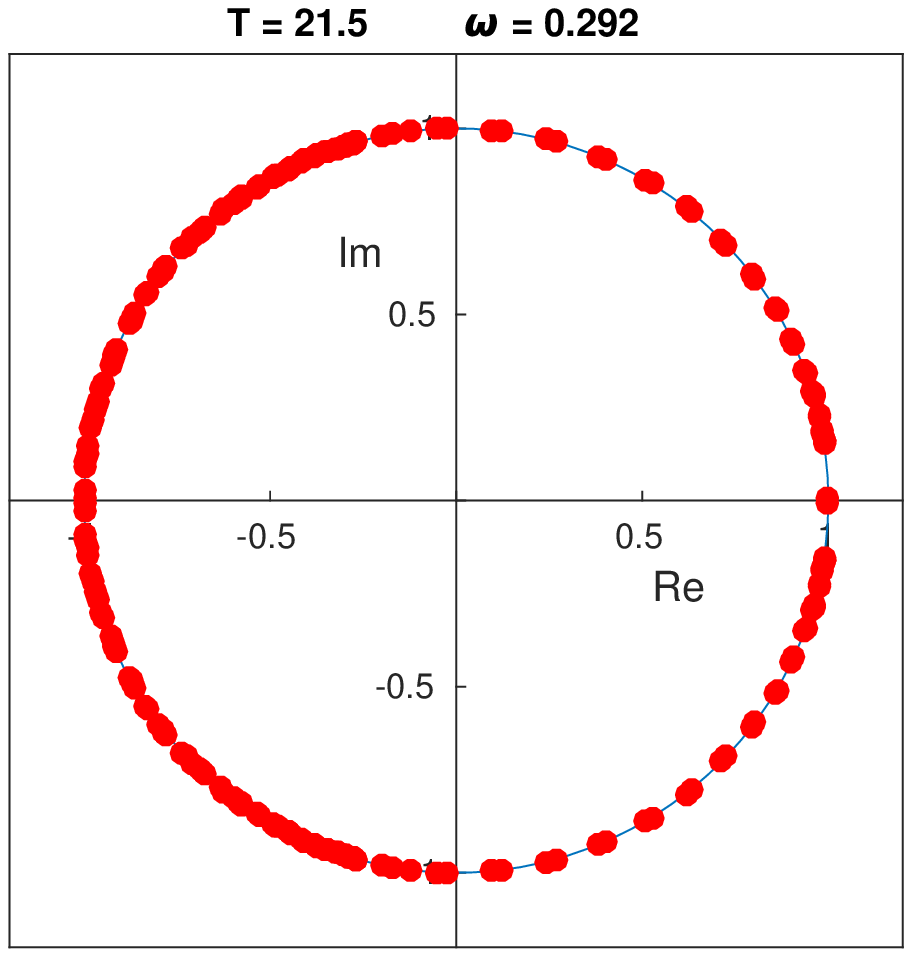}
\\
\end{tabular}
\caption{The intersite family for $T=16.7, 17, 18.5\ \text{and}\ 21.5$ (or equivalently $\w=0.376, 0.370, 0.340\ \text{and}\ 0.292$), together with 
the corresponding Floquet multipiers. We can see here that this family is linearly 
(spectrally) stable, since all multipliers stay on the unit circle for 
all values of $\w$ in this interval.}
\label{intersite}
\end{figure}

The inter-site family for $T=16.7-21.5$ (or $\w=0.376-0.292$) is shown in Fig.~\ref{intersite}. 
This family is linearly stable {\it throughout} the relevant frequency range, since 
all Floquet multipliers lie on the unit circle for all values of $\w$.

\subsection{Bifurcation from the linear limit}

\begin{figure}
\centering
\includegraphics[scale=0.6]{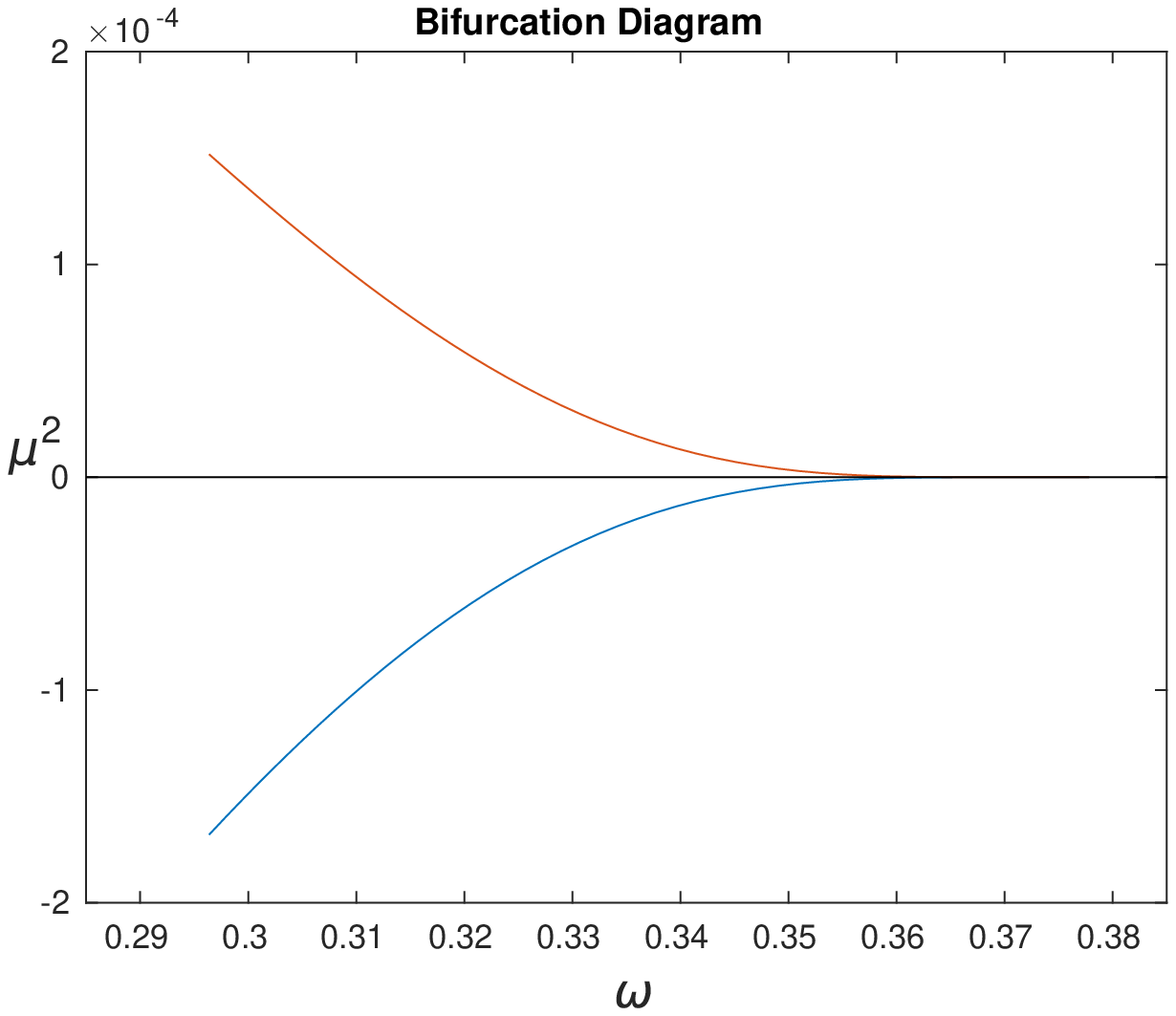}
\caption{The dependence of the square of the appropriate
characteristic exponent $\mu$ of the onsite and inter-site families (see also
the discussion in the text), with respect to the frequency of the breather. 
In the upper half-plane, the value of the square of the largest real characteristic 
exponent is shown for the onsite family. In the lower half-plane, the square 
of the value of one of the two isolated characteristic exponents which lie close to 1
(yet on the unit circle), is shown. Since this exponent is purely imaginary, 
$\mu^2$ is negative. The former characterizes the instability
of the onsite branch, while the latter illustrates the spectral
stability of the inter-site branch.}
\label{bifdiag}
\end{figure}

Both the onsite and inter-site families bifurcate from the same
linear mode at the band edge of $k=\pi$.
This can be seen in two different ways. In Fig.~\ref{bifdiag}, the
dependence of the square of an appropriate characteristic exponent $\mu$ of the onsite and inter-site 
families, with respect to the frequency of the breather is shown. By the term appropriate we refer to one of the exponents of the corresponding family which is indicative of the stability of the solutions. We remind here that the relationship between the characteristic exponents and the corresponding Floquet 
multipliers is given by $\lambda=e^{\mu T}$. So, for a solution to be linearly unstable it is enough to have a pair of real exponents while in order to have a linearly stable solution it is required all of the exponents to lie on the imaginary axis.
Thus, in order to demonstrate the stability of the breather families, in the upper half-plane of this diagram, 
the value of the square of the largest real characteristic exponent is shown for the onsite family. 
In the lower half-plane of the diagram, the square of the value of one of the two 
isolated characteristic exponents which lie close to 0 (but on the imaginary axis), is depicted. Since this 
exponent is purely imaginary, $\mu^2$ is negative. This confirms the 
instability of the former and the (spectral) stability of the latter family. 

\begin{figure}
\centering
\includegraphics[scale=.38]{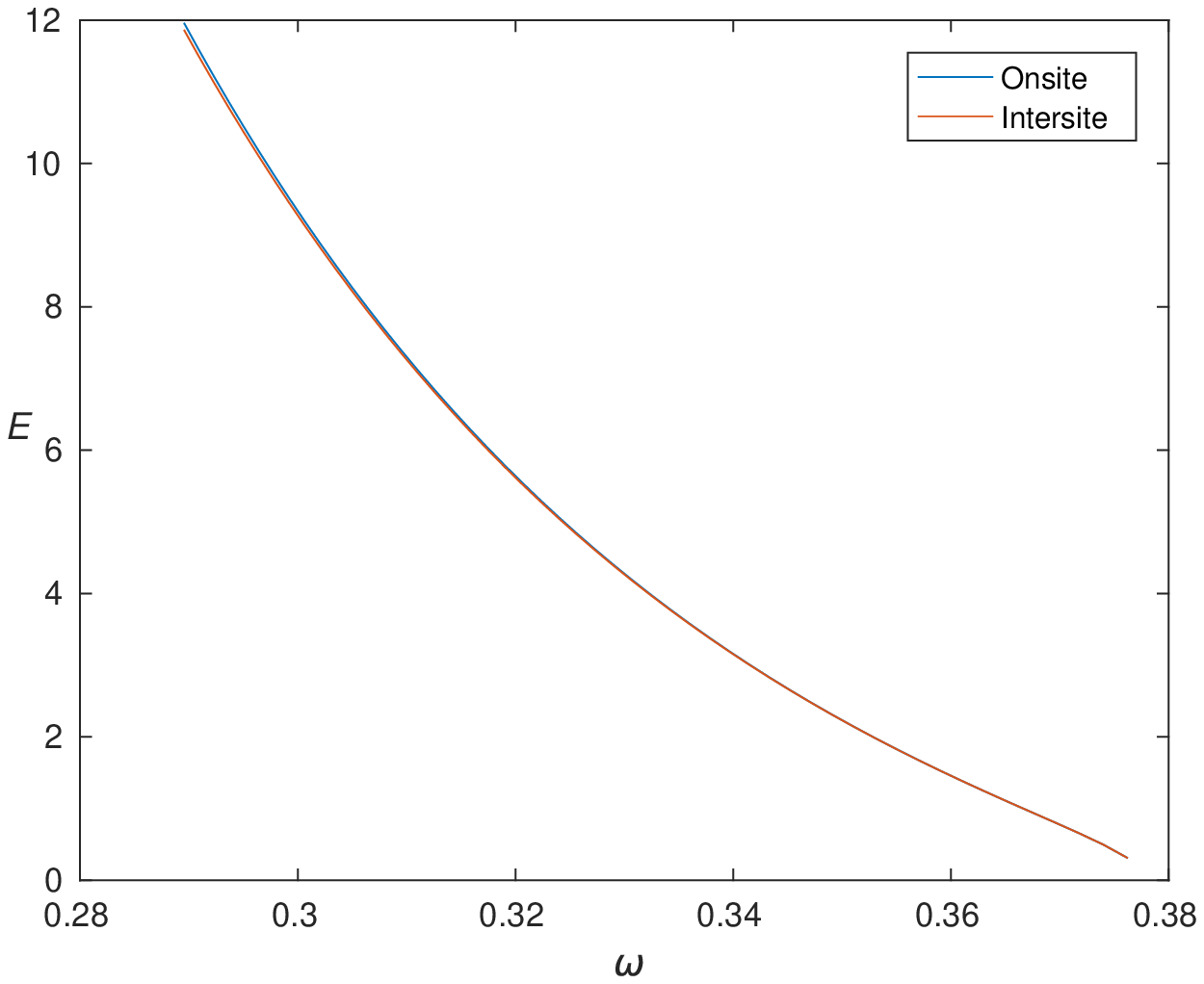} 
\includegraphics[scale=.38]{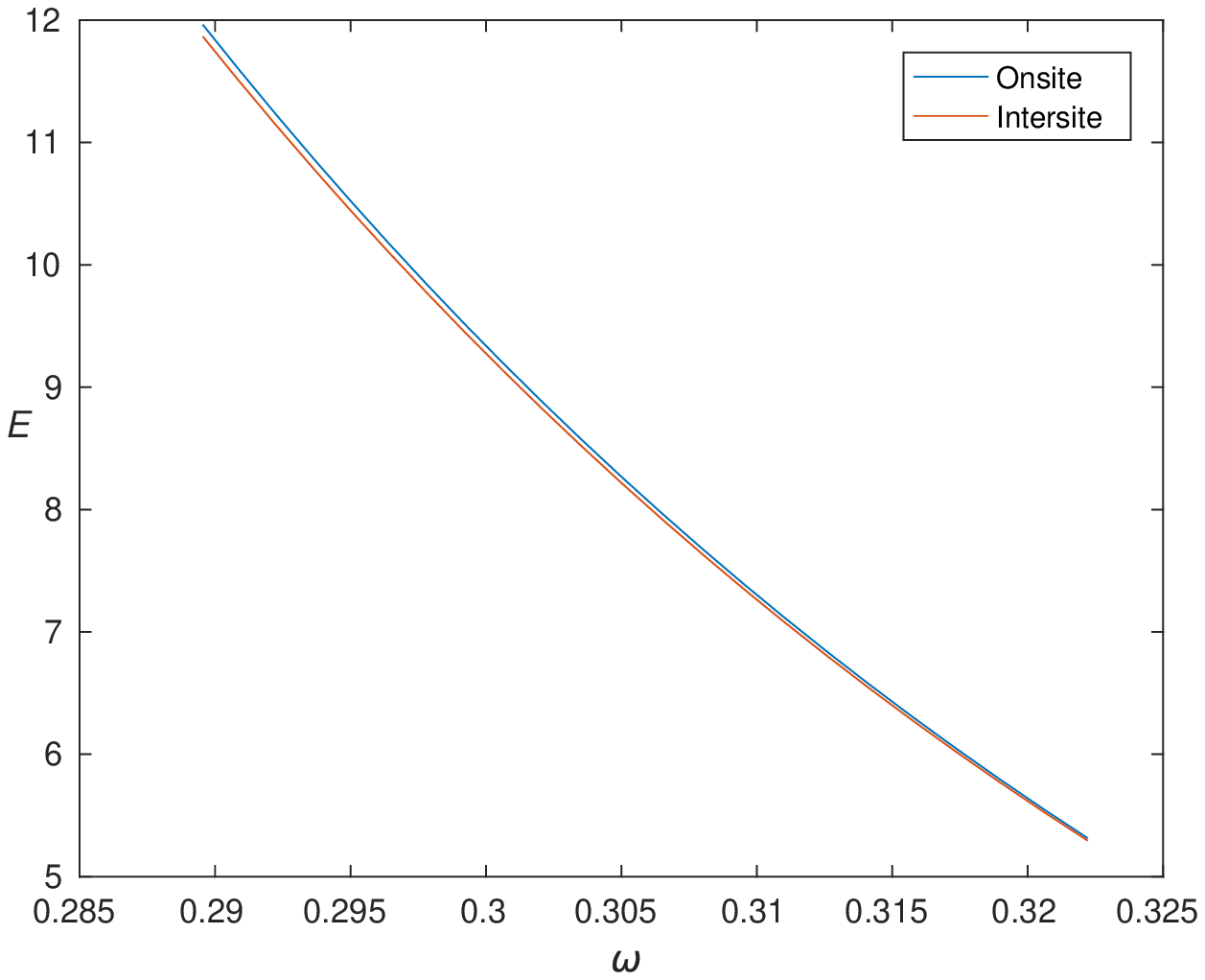}
\includegraphics[scale=.38]{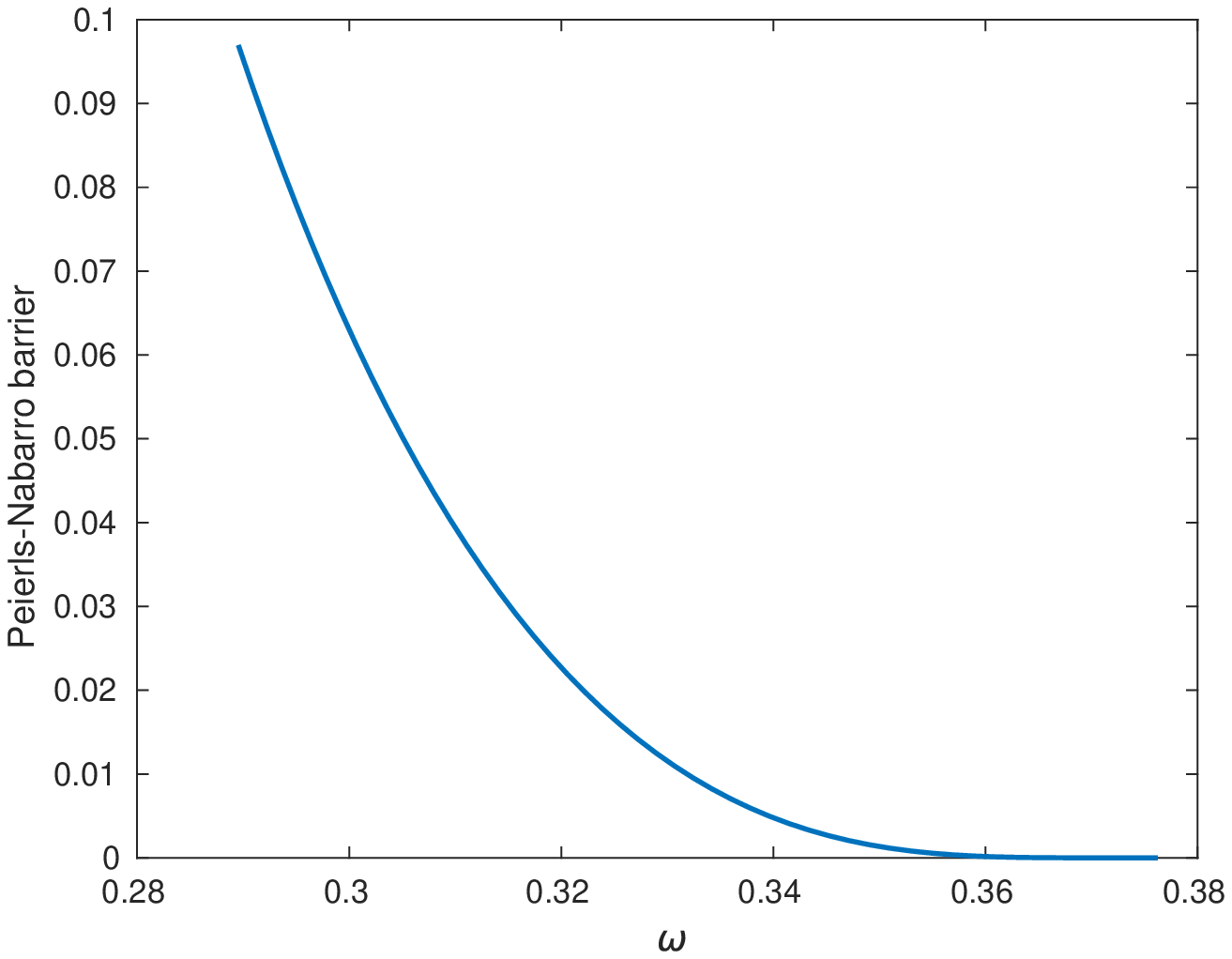} 
\caption{Left and middle panel: the bifurcation diagram of the total energy of the lattice with respect to the frequency of the breather.
The two curves almost coincide. The lower energy
curve corresponds to the
inter-site solution (local energy minimum)
while the higher to the onsite one (saddle
point of the energy surface).
This fact is energetically consistent with the spectral stability
of the former solution and the instability of the latter.
In the left panel the curves for the whole range of existence of the breather are shown. In the middle panel a sub-interval
of the frequency range is considered in order to more lucidly
illustrate the
difference between the two curves.
The right panel presents the difference between the onsite energy and inter-site energy curves
(the PN barrier) as a function of the frequency.
}
\label{bifdiag2}
\end{figure}

A second way to visualize this bifurcation is to depict the total
energy of the lattice (measured in accordance with Eq.~(\ref{Etot})) 
for the two distinct configurations with respect to the frequency of the solution.
In Fig.~\ref{bifdiag2} the two resulting curves are shown. 
The two curves almost coincide and their difference is rather difficult to discern.
There is a small divergence in the left part of the diagram. 
The lower curve corresponds to the inter-site solution. 
This fact is energetically consistent since the lower energy corresponds 
to the linearly stable solution (a local energy minimum). The central panel showcases
a sub-interval of the entire frequency interval in order to 
clarify the energy difference. Finally, the right panel 
shows the energy difference as a function of $\omega$, 
presenting the rapid growth of this quantity as a function 
of the frequency. The significant dynamical implications 
of this rapid growth will be explored in the next subsection. 

\subsection{Breather Dynamics}

Let us now examine the dynamical behavior of the linearly unstable
(onsite) breather solutions under perturbation. For the inter-site
waveforms, we have simply confirmed their dynamical stability;
in this case, upon weak perturbations, nothing interesting
happens. The perturbation we consider for the unstable 
onsite solutions is one along the direction of the eigenstate, 
which corresponds to the largest unstable Floquet multiplier of
the $T=19$, $\w=0.331$ onsite breather multiplied by a prefactor $0.1$.
Arguably, there is no single optimal way to perform the 
relevant perturbation. Here, we opted to perturb all the 
members of the breather family by the same ``kick'' (however, even 
in that case, the kick does not have same projection to the 
respective unstable eigendirection of each family member). 

\begin{figure}
\centering
\includegraphics[scale=0.5]{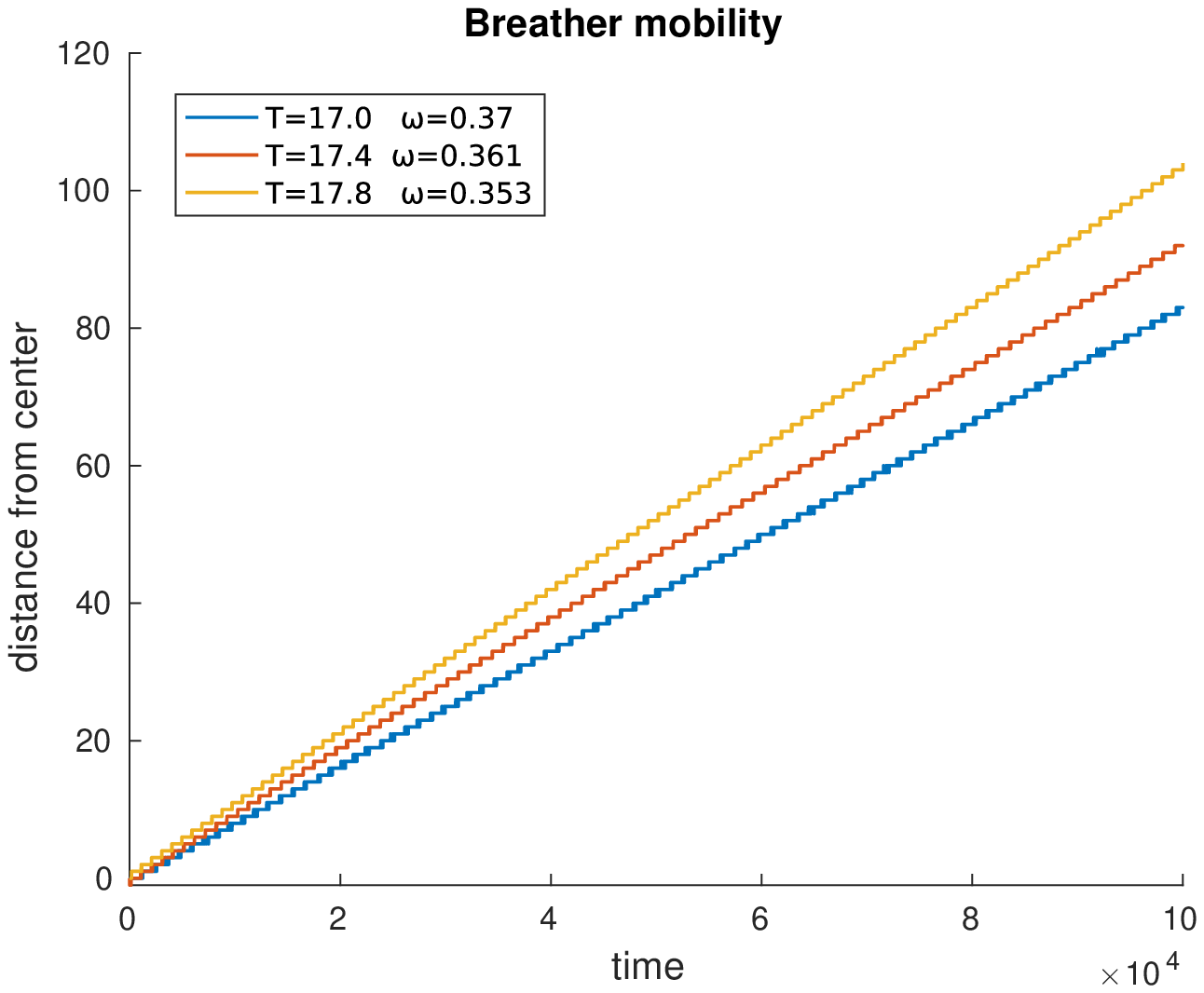}
\includegraphics[scale=0.5]{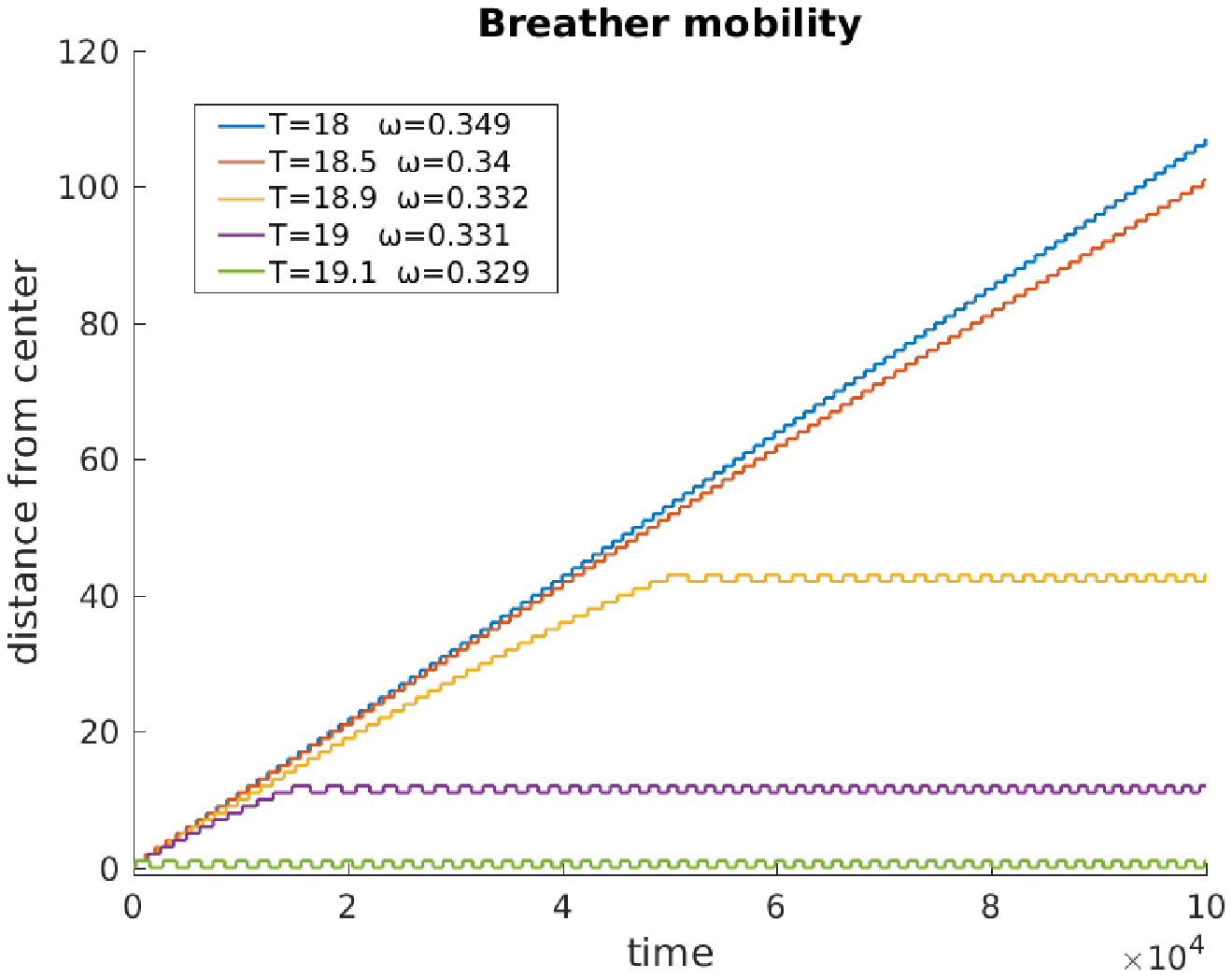}
\caption{(Color Online) Breather mobility under perturbation. 
Left panel: The mobility of the breather increases as the frequency decreases. 
Right panel: The mobility of the breather decreases as the frequency decreases.}
\label{brmob}
\end{figure}

The results of the dynamical simulations are shown in Fig.~\ref{brmob}, 
where the position of the maximum of the energy-per-site (defined in Eq.~(\ref{En})) is depicted with 
respect to the time elapsed. In general we observe that under this perturbation
the breathers exhibit some mobility. 
One can think of this as increasing the energy of the saddle point, 
(by perturbing it along the unstable eigendirection) 
giving it a kick that allows it to become mobile 
in the direction connecting the onsite energy maxima with the
inter-site energy minima.
In the left panel the behavior of the breathers with $T=17 - 17.4$ is
depicted. In this case as we move away from the bifurcation point i.e. for increasing values of the period 
(decreasing values of the frequency) the mobility of the breather is increased. This happens because as the 
period is increased more energy is channeled towards the mobility direction since the perturbation used is the 
one which corresponds to the period value $T=19$. As the values of the period (frequency) are further 
increased (decreased), we observe the opposite phenomenon
(right panel of Fig.~\ref{brmob}), i.e., the breather mobility is decreased. 
In this case, we are close to the period of which the eigenstate we use for the perturbation. As the period 
increases further (frequency decreases), the Peierls-Nabarro barrier becomes larger (rapidly so), 
so the mobility decreases. In particular, for periods $T=18,~18.5$ ($\w=0.349,~0.34$), 
we observe a linear increase of the distance of the breather from the center, 
while for periods $T=18.9, ~19$ ($\w=0.332,~0.331$) the breather moves for some 
lattice sites and then it stops. Finally, for $T=19.1$ ($\w=0.329$), and all the 
larger period (smaller frequency) values, the breather does not move at all. 
Instead, it performs a small oscillation around the originating lattice site.     
This is strongly reminiscent of the dramatically different (in
comparison to the continuum) highly discrete behavior originally identified in models such
as the sine-Gordon and other Klein-Gordon ones~\cite{pk84,memiw}, 
and also explored in other such models, including the discrete
nonlinear Schr{\"o}dinger ones~\cite{jenkinson}.

\begin{figure}
\centering
\includegraphics[scale=0.38]{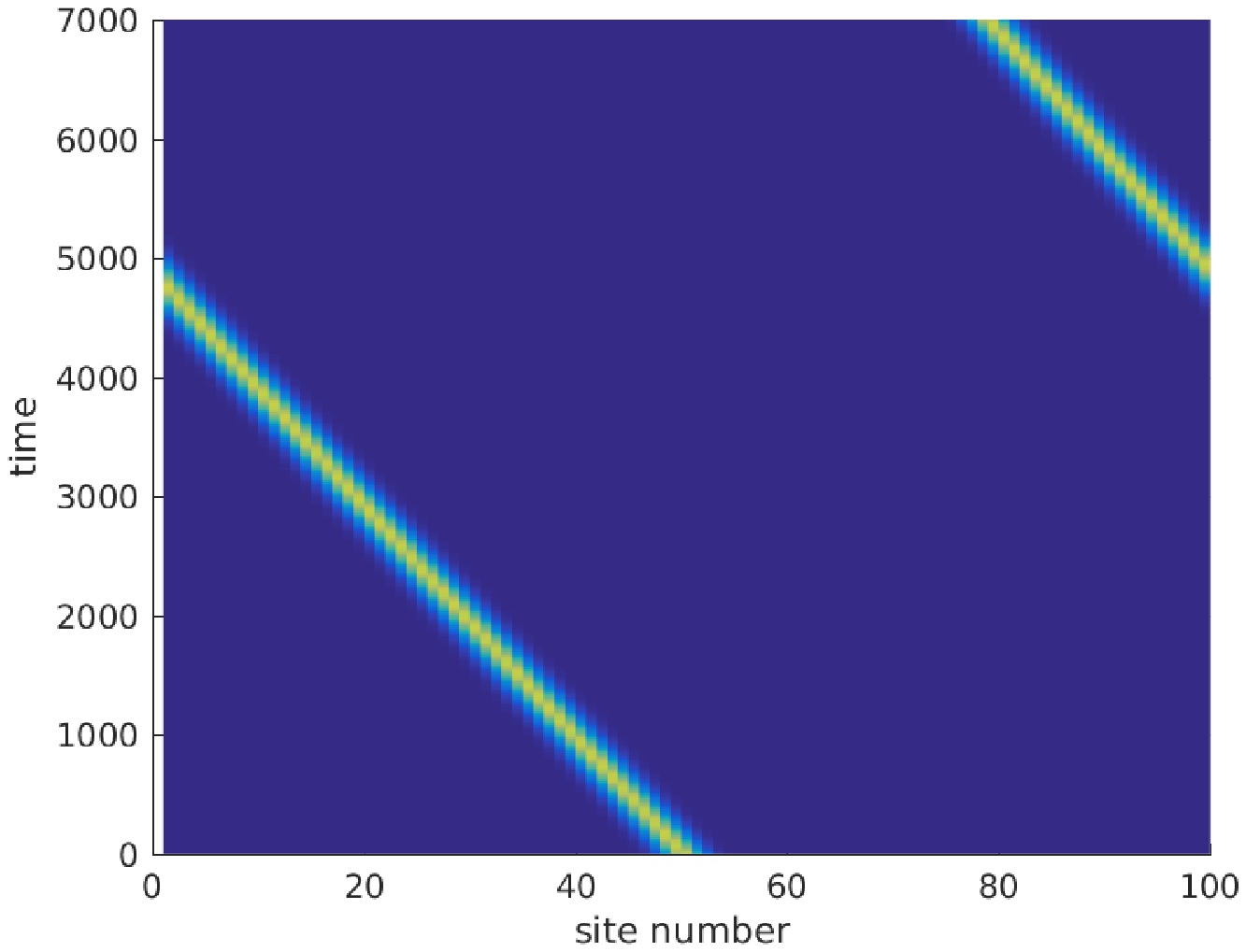}
\includegraphics[scale=0.38]{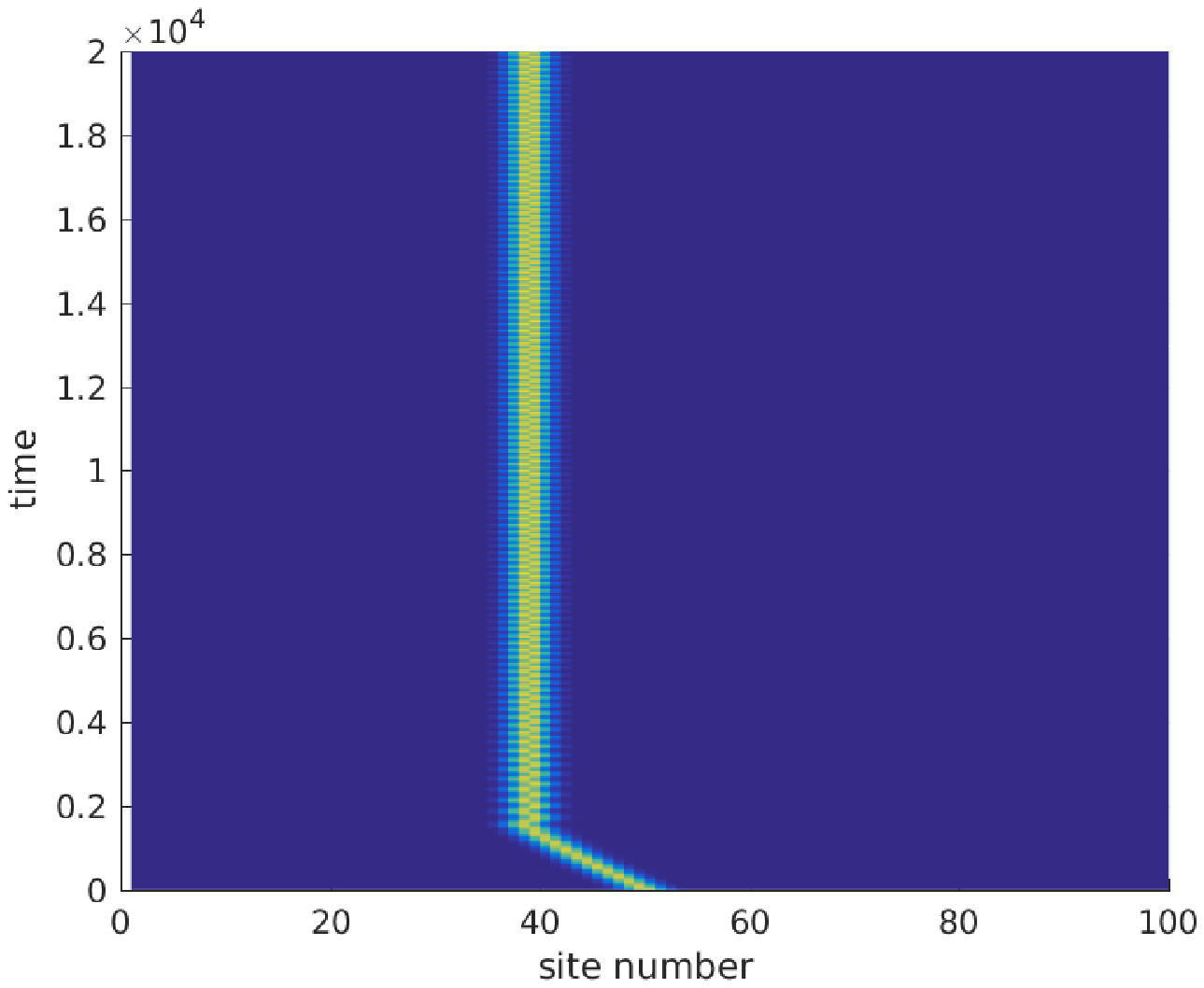}
\includegraphics[scale=0.38]{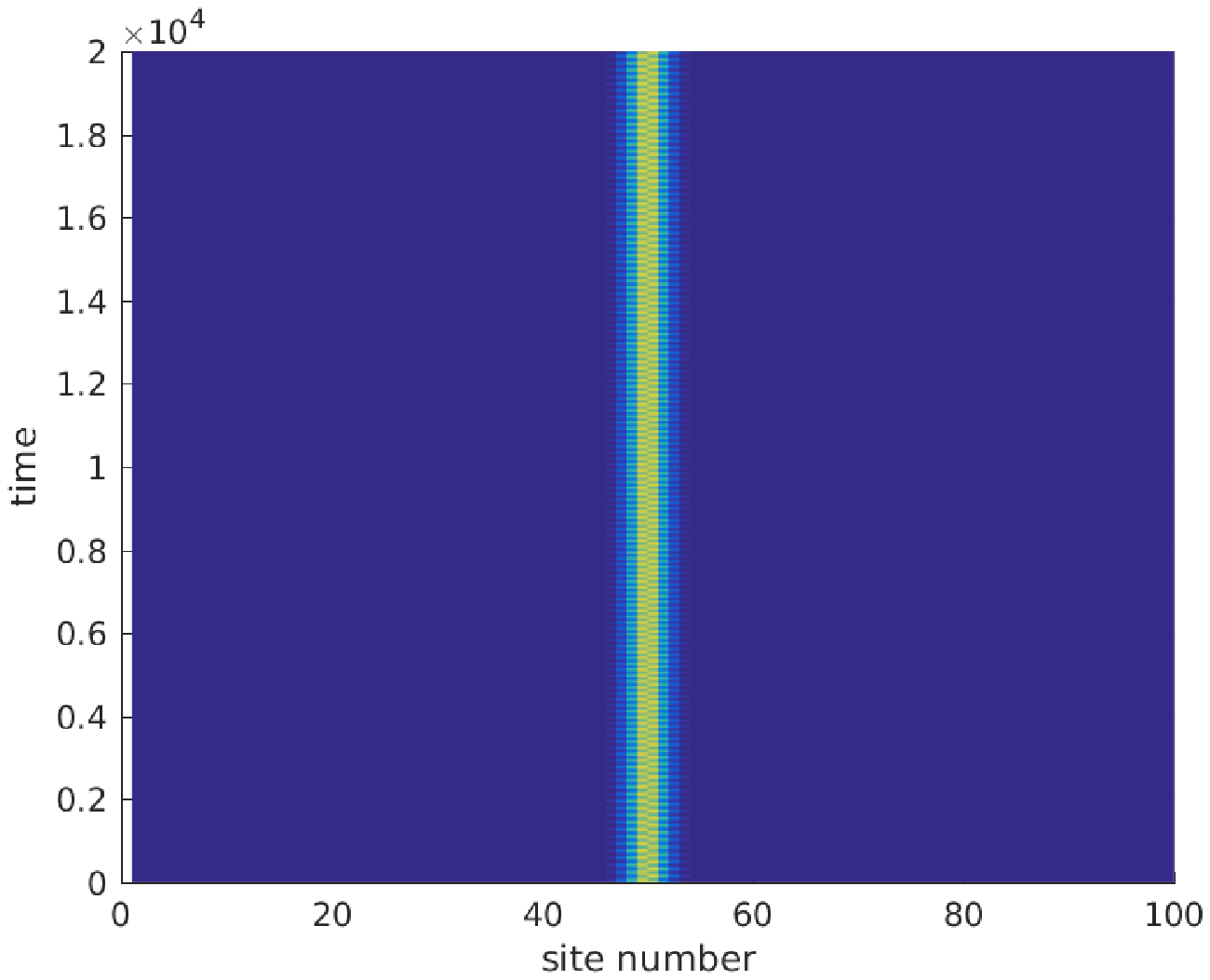}
\caption{(Color Online) Contour plots of the energy-per-site 
with respect to the site number and time are shown. 
In the left panel, the linear motion of the breather is shown for the case $T=18.5,~\w=0.34$.
The breather passes to the opposite side of the graph because we have considered periodic boundary conditions. 
Here, the breather energy can be freely transferred within the lattice.
In the central panel, the case $T=19,~\w=0.331$ is considered. 
The breather moves for some lattice sites and then it stops,
being trapped and performing small oscillations around a single
local energy minimum (a corresponding inter-site configuration).
In the right panel for the $T=19.3,~\w=0.326$ breather, under the specific perturbation, 
it is never possible for the breather to traverse the lattice. Instead, 
it performs only small oscillations around the corresponding local energy minimum 
intersite configuration.}
\label{contourbrmob}
\end{figure}

In Fig.~\ref{contourbrmob}, the contour plot of the energy-per-site with 
respect to the site number and time is shown. This gives a 
transparent picture of the distribution of the energy, 
illustrating the transition from high mobility
(left panel) to partial mobility (middle panel) and finally
to complete trapping and pinning between two lattice
sites (right panel). It can be clearly inferred that in the
vicinity of the linear limit (the limit where the NLS
and multiscale description are accurate), the breather
is highly mobile. As the frequency is decreased, the
dynamics becomes progressively more discrete and
the breathers less mobile.

\section{Conclusions and Future Challenges}

In the present work, we have systematically examined the
possibility of left-handed metamaterial lattices to support
discrete breathers. Starting from the underlying nonlinear
transmission line model, and unveiling its energy preserving
structure, we developed a nonlinear Schr{\"o}dinger-based
multiscale analysis. This formulation operated as a yardstick
that revealed both the underlying linear structure, but importantly
also revealed the nature of the weakly nonlinear limit. More
specifically, it rendered evident that among the two stationary
breather limits, the one that was promising for bright breathers
was that of $k=\pi$ (i.e., of anti-symmetric nearest neighbor
excitations). There, the bright soliton ansatz of the NLS
could be used to reconstruct approximate bright breathers
of the original (transmission line) lattice model. 

Contrary to earlier works simply testing in direct numerical simulations
the robustness of such states, here we went a significant step 
further. We identified the numerically exact solutions 
and continued them parametrically over the frequency.
We revealed that similarly to other models of both kink
and breather-bearing type~\cite{pk84,memiw,jenkinson},
the bright breathers herein come in two varieties,
namely onsite and inter-site. The detailed numerical
linear stability thereof was used to establish that
the former solution is unstable and the latter one is
stable. Finally, the relevant solutions were used
in dynamical evolution numerical experiments confirming
that the unstable onsite configurations could, in principle,
become mobile. However, this mobility was far more significant
for frequencies near the linear limit and was considerably
more restricted, as the solutions became more highly
nonlinear, and more narrow (far from the linear limit).

This study paves the way for numerous related explorations
in the future. We considered here a model where the capacitance
increases with the voltage. If instead,
the capacitance decreases with the voltage (as is,
e.g., the case in many nonlinear ceramic capacitors), it
may be natural to expect that dark breathers may arise,
instead of bright ones. The latter are considerably
more technically involved to compute (see, e.g., Ref.~\cite{chong}
and discussion therein) chiefly due to the finite background
of the solutions and thus pose a natural challenge
towards future work. Additionally, the waveforms examined 
here, as well as in the earlier work of Ref.~\cite{inter}
have been restricted to one-dimensional settings.
Extending consideration to two-dimensional scenarios
where the geometry may also play a role (e.g., hexagonal
or honeycomb lattices, rather than purely square ones)
is then another relevant possibility, especially given recent
experimental realizations~\cite{lars2d}. Finally,
the consideration of composite lattices bearing
both right-handed and left-handed elements (see
for an example~\cite{vel}) is another topic of interest
in its own right as concerns the nonlinear (numerically
exact) breather waveforms that may arise therein.
These variants of the present problem are currently under
consideration and will be reported in future studies.

{\bf Acknowledgments.} The authors, V.K., P.G.K., G.P.V., and D.J.F.,  
acknowledge that this work made possible by NPRP grant {\#} [9-329-1-067] 
from Qatar National Research Fund (a member of Qatar Foundation). 
The findings achieved herein are solely the responsibility of the authors.
We also thank Professor Yannan Shen for numerous useful discussions. V.K. and P.G.K. are  grateful for support  from the
ERC under FP7, Marie Curie Actions, People, International Research Staff Exchange Scheme (IRSES-606096).

\end{document}